\documentclass[12pt]{article}
\usepackage[utf8]{inputenc}
\usepackage[svgnames]{xcolor}
\usepackage[a4paper, inner=1cm, outer=1cm, top=1cm, bottom=2cm]{geometry}
\usepackage[bottom, multiple]{footmisc}
\usepackage{graphicx}
\usepackage{tikz}
\usetikzlibrary{decorations.markings}
\usetikzlibrary{decorations.pathmorphing}
\usepackage{framed}
\usepackage{csquotes}
\definecolor{shadecolor}{rgb}{0.90,0.90,0.90}
\usepackage{cite}
\usepackage{hyperref}
\usepackage{subcaption}
\usepackage{pifont}
\usepackage{dingbat}
\usepackage{setspace}
\usepackage{amsmath, amssymb, amsthm, float, graphicx,amsfonts,braket}
\numberwithin{equation}{section}
\usepackage{amsthm}
\newtheorem{theorem}{Theorem}[section]
\theoremstyle{definition}

\hypersetup{colorlinks=true, linkcolor=Maroon, citecolor=FireBrick,urlcolor=Green,linktocpage}

\def\beq{\begin{eqnarray}}\def\eeq{\end{eqnarray}}
\def\be{\begin{equation}}\def\ee{\end{equation}}

\def\p{\pi}

\def\s{\sigma}
\def\m{\mu}

\def\a{\alpha}
\def\e{\epsilon}
\def\ve{\varepsilon}

\def\d{\delta}

\def\t{\tau}
\def\D{\Delta}

\def\ma{{\mathcal{A}}}
\def\md{{\mathcal{D}}}
\def\me{{\mathcal{E}}}

\def\mm{{\mathcal{M}}}

\def\mW{{\mathcal{W}}}
\def\nn{\nonumber}

\usepackage{bm}
\usepackage{bm}

\begin{document}

\title{\bf Locality and Analyticity of the Crossing Symmetric Dispersion Relation}
\date{}
\author{Debapriyo Chowdhury$^{\a}$\footnote{debapriyoc@iisc.ac.in}, Parthiv Haldar$^{\a}$\footnote{parthivh@iisc.ac.in}, and Ahmadullah Zahed$^{\a}$\footnote{ahmadullah@iisc.ac.in
}\\~~~~\\
\it ${^\a}$Centre for High Energy Physics,
\it Indian Institute of Science,\\ \it C.V. Raman Avenue, Bangalore 560012, India.}
\maketitle
\vskip 2cm
\maketitle
\vskip 2cm
\abstract{This paper discusses the locality and analyticity of the crossing symmetric dispersion relation (CSDR). Imposing locality constraints on the CSDR gives rise to a local and fully crossing symmetric expansion of scattering amplitudes, dubbed as Feynman block expansion. A general formula is provided for the contact terms that emerge from the expansion. The analyticity domain of the expansion is also derived analogously to the Lehmann-Martin ellipse. Our observation of type-II super-string tree amplitude suggests that the Feynman block expansion has a bigger analyticity domain and better convergence. 
}

\tableofcontents

\onehalfspacing

\section{Introduction}
Dispersion relations are non-perturbative techniques for scattering amplitudes. The usual practice of writing dispersion relations for 2-2 scattering of identical particles involves keeping the Mandelstam invariant $t$-fixed and writing down a Cauchy integral in the $s$-variable, leading to an $s \leftrightarrow u$ symmetric representation of the amplitude. Then crossing symmetry is imposed as an additional condition, which gives rise to the null constraints. There are many recent advancements in constraining quantum field theories using dispersion relations, which rely on certain analyticity and unitarity assumptions \cite{Dispcause, Martinlec, Froissartbnd, Meltzer, nima, Mizera}. For example, the constraints on EFT Wilson coefficients were worked out in \cite{RMTZ,rattazzi,TWZ,Wang:2020jxr,chout,chout2, chout3, Vichi,nimayutin,green,nimayutin2,yutin2,sasha,Davis:2021oce}. Dispersion relations and their connection to geometric function theory was discussed in \cite{PHASAZ, PRAS,AZ}. A fascinating connection between dispersion relations and knots was discussed in \cite{AS}. Similar techniques have also been developed for CFT Mellin amplitudes, see for example \cite{joaopaper,joaopaper2,cmrs,RGASAZ}.

Even though fixed-$t$ dispersion has had enormous success, it lacks full crossing symmetry. As we know from the Feynman diagram calculations, crossing symmetry is manifest. To achieve a crossing symmetric version of the dispersion relation, one must write down parametric dispersion relations by parametrizing the Mandelstam invariants $s,t$ as functions of crossing symmetric variables $z,a$. After this, we can write down the Cauchy integral in $z$-variable for fixed-$a$. This dispersion relation will be fully crossing symmetric. In $s,t$ variables or the shifted Mandelstam variables $s_1=s-\m/3,\,s_2=t-\m/3,\,s_3=u-\m/3,~ \m=4m^2$, one obtains the  \emph{crossing-symmetric dispersion relation (CSDR)} \cite{AK, ASAZ}
\be\label{dis1} 
\mm( s_1, s_2)=\a_0+\int_{\frac{2\m}{3}}^\infty \, \frac{d\s}{\s}\, \ma(\s;\t^+(\s,a))\,H(\s;\, s_1, s_2,s_3),
\ee where $\ma(s_1,s_2)$ is the $s-$channel absorptive part of the amplitude, the discontinuity of the amplitude across the cut $s_1\ge\,\frac{2\m}{3}$, $\a_0=\mm(s_1=0, s_2=0)$ is the subtraction constant, and  
\begin{align}
	\t^+(\s,a)&:=-\frac{\s}{2}\left[1-\left(\frac{\s+3a}{\s-a}\right)^{\frac{1}{2}}\right],\\
H(\s;\, s_1,s_2,s_3)&:=\frac{s_1}{\s-s_1}+\frac{s_2}{\s-s_2}+\frac{s_3}{\s-s_3}, \quad s_1+s_2+s_3=0.
\end{align}
The variable $a$, which is central to our analysis, is a crossing symmetric variable defined by 
\be \label{adefintro}
a= y/x,\qquad x=-(s_1s_2+s_2s_3+s_3s_1),\,\, y=-s_1s_2s_3.
\ee Further developments and applications can be found in \cite{roy,Roskies:1970uj,anant}. 

A fully crossing symmetric local scattering amplitude can be Taylor expanded in the elementary crossing symmetric polynomials $x, y$ as following
\be 
\mm( s_1, s_2)=\sum_{p=0,q=0}^{\infty} \mW_{p,q} x^p y^q, 
\ee
$\mW_{p,q}$ being the Wilson coefficients.   {Locality requires the power series expansion above only comprising non-negative powers of $x,y$. However if we consider just the dispersive representation \eqref{dis1} for the amplitude $\mm( s_1, s_2)$, there is no obstruction to negative powers of $x$. As mentioned before, locality requires these terms to be absent. This obstruction  realizes itself into non-trivial constraints on the partial wave coefficients of $\ma(s_1,s_2)$, known as \emph{locality constraints} \cite{ASAZ}, which can be recast into the \emph{null constraints} \cite{chout} that one gets from the fixed-$t$ dispersion relation after imposing crossing as an additional constraint.
 
 The crossing-symmetric dispersion relation  along with locality constraints provides a new kind of \enquote{block expansion} for the amplitude which directly relates with usual Feynman diagrammatic expansion of perturbative amplitude. If we insert the usual partial wave expansion in terms of the Gegenbauer polynomials $C_\ell^{(\a)}, \a=(d-3)/2,$ of the absorptive part \be
\ma(s_1, s_2)=\Phi(s_1;\a)\, \sum_{\ell}(2\ell+2\a)\,a_\ell(s_1)\, C_\ell^{(\a)}\left(1+2 \frac{s_{2}+\frac{\mu}{3}}{s_{1}-\frac{2 \mu}{3}}\right),
 \ee $a_\ell(s_1)$ being the partial wave coefficients of the absorptive part, and $\Phi(s_1;\a)=\Psi(\a) \sqrt{s_1+\m/3}/(s_1-2\m/3)^\a$, $\Psi(\a)$ being a positive number depending only on $\a$,    into the CSDR then we get a ``block" expansion called \emph{Dyson block expansion} \cite{ASAZ}. The Dyson blocks $\{M_\ell^{(D)}(\s; s_1, s_2, s_3)\}$ are defined as 
 \be 
 M_\ell^{(D)}(\s; s_1, s_2, s_3)= Q_\ell(s_1, s_2)\, H(\s; s_1, s_2, s_3),\quad Q_{\ell}\left(s_{1}, s_{2}\right)=\left(s_{1}-\frac{2 \mu}{3}\right)^{\ell} C_{\ell}^{(\alpha)}\left(1+2 \frac{s_{2}+\frac{\mu}{3}}{s_{1}-\frac{2 \mu}{3}}\right).
 \ee Dyson blocks contain 
 removable non-local terms. These non-local terms can be removed systematically. If we remove these non-local terms then, we are left with a block expansion which is manifestly crossing symmetric \emph{and} local. This expansion is called \emph{Feynman block} expansion \cite{ASAZ}. The Feynman block expansion is so named because this makes direct connection with Feynman diagrams as explained in \cite{ASAZ}. This block expansion reads as follows:
\be
\begin{aligned}
	&\mathcal{M}\left(s_{1}, s_{2}\right)=\alpha_{0}+\frac{1}{\pi} \sum_{\ell=0}^{\infty}(2 \ell+2 \alpha)\, \int_{\frac{2 \mu}{3}}^{\infty}\, d \sigma\,  \frac{\Phi(\sigma ; \alpha)}{\left(\sigma-\frac{2 \mu}{3}\right)^{\ell}} \,  a_{\ell}(\sigma)\, M_{\ell}^{(F)}\left(\sigma ; s_{1}, s_{2}\right), 
\end{aligned}
\ee

The Feynman blocks $\{M_\ell^{(F)}(\s;s_1, s_2)\}$ are defined by 
 \be
M_{\ell}^{(F)}\left(\sigma ; s_{1}, s_{2}\right):=\sum_{i=1}^{3} M_{\ell}^{(i)}\left(\sigma ; s_{1}, s_{2}\right)+M_{\ell}^{(c)}\left(\sigma ; s_{1}, s_{2}\right),
\ee
where
\be
\begin{split}
	&M_{\ell}^{(1)}\left(\sigma ; s_{1}, s_{2}\right)=Q_{\ell}\left(s_{1}, s_{2}\right)\left(\frac{1}{\sigma-s_{1}}-\frac{1}{\sigma}\right) \,,\\
	&M_{\ell}^{(2)}\left(\sigma ; s_{1}, s_{2}\right)= M_{\ell}^{(1)}\left(\sigma ; s_{2}, s_{3}\right), \quad M_{\ell}^{(3)}\left(\sigma ; s_{1}, s_{2}\right) = M_{\ell}^{(1)}\left(\sigma ; s_{3}, s_{1}\right),
\end{split}
\ee
and $M_\ell^{(c)}(\s;s_1,s_2)$ are the  \emph{contact terms} containing terms  \emph{polynomial} in $\{s_i\}$.}
We have derived a general formula for the \emph{contact terms} 
\be\label{introcont}
\boxed{M_{\ell}^{(c)}\left(\sigma ; s_{1}, s_{2}\right)=\sum _{p=0}^{\ell } \sum _{q=0}^{\ell } \frac{b_{p,q}^{(\ell)}}{2}\left[-\mathcal{J}_{\frac{\ell }{2},p,q}^{(F)}(\sigma,x,a)+\sum _{n=1}^{\frac{\ell }{2}} \sum _{m=0}^n \mathcal{J}_{n,m,p,q}^{(D)}(\sigma,x,a)\right],}
\ee
where $a=y/x, b_{p,q}^{(\ell)}=\frac{1}{p!\,q!}\partial_{s_1}^p\partial_{s_2}^q\,Q_\ell(0,0)$ and $\mathcal{J}_{\frac{\ell }{2},p,q}^{(F)}(\sigma,x,a),\mathcal{J}_{n,m,p,q}^{(D)}(\sigma,x,a) $ are known functions, whose expressions have been derived in the main text, \eqref{JFM}, \eqref{JDM}. This expression is one of the main results obtained in the present paper. 

\begin{figure}[hbt!]
\centering
\includegraphics[scale=0.35]{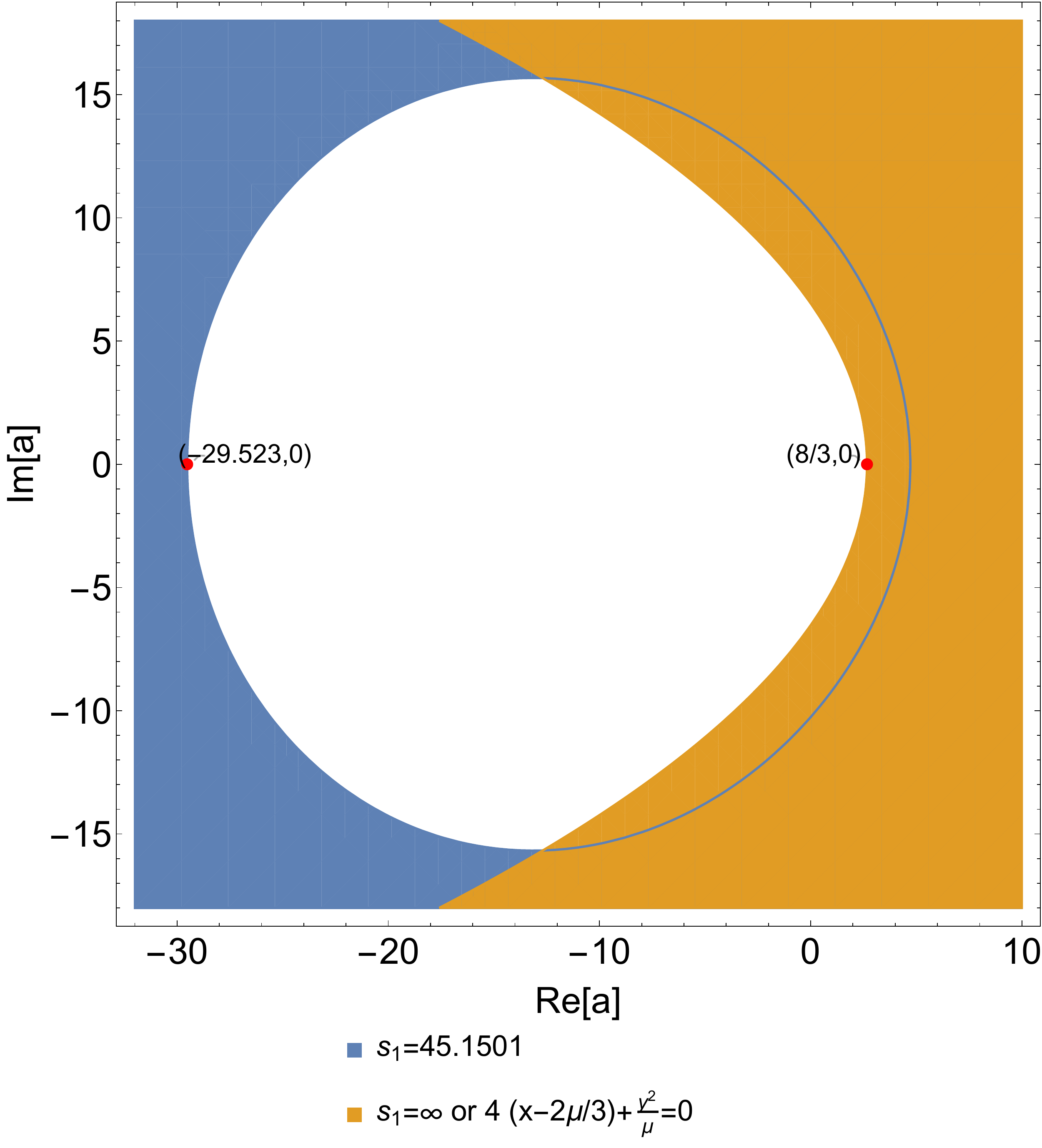}
\caption{The analyticity domain in the complex $a$-plane for $\frac{2 \mu }{3}<s_1<\infty$. Shaded regions are excluded.}
\label{fig:allrange1}
\end{figure}

In the second part of the work, we investigate the analyticity domain for the dispersion relation \eqref{dis1} in complex $a$. Our starting point is the analogue of the Lehmann-Martin ellipses. We numerically investigate the convergence of Dyson and Feynman block expansions for specific S-matrices.  The analyticity domain in the complex $a$-plane for physical $s_1\in \left( \frac{2 \mu }{3}, \infty\right)$  is shown in the  figure \eqref{fig:allrange1}. The white region is the domain that we find out, and various numerical checks support our analyticity domain of the Feynman block.

{Let us tabulate the main results that we have obtained in the present work. 
\begin{itemize}
    \item We have derived an analytic expression of the Feynman blocks introduced in \cite{ASAZ}. The non-trivial part of the Feynman block is the contact terms, the expression of which, \eqref{introcont}, has been derived.
    
	\item {We derived the domains of the analyticity of the scattering amplitude in the crossing-symmetric dispersive representation, namely the Dyson and Feynman block expansions in complex $a$.}
    
    \item  We provide numerical support to our argument in specific amplitudes, $4$- dilaton scattering amplitude in tree level type-II superstring theory, and Pion S-matrices. We have found evidence that Feynman block expansion converges in domain larger than that of the Dyson block expansion.   
\end{itemize}
}
\vspace{0.5cm}
The rest of the paper is organized as following. Section $2$ reviews crossing symmetric dispersion relations and sets up the notations. Section $3$ gives detailed definitions of the Dyson block and Feynman block and derives the general formula for contact terms. Section $4$ reviews the convergence of partial wave expansion of scattering amplitudes and Lehmann-Martin ellipses. The analyticity domain of the crossing symmetric dispersion relations is derived and various illustrations are presented in Section 5. Section 6 gives the analyticity domain of the Dyson and Feynman block. In Section $7$, we summarize and conclude with future directions.

\section{Crossing symmetric dispersion relation}
The premise of our analysis is the dispersive representation presented in \cite{AK, ASAZ}, called crossing-symmetric dispersion relation (CSDR), of the scattering amplitude. We give a quick review of the same to set the stage. One of the problem of fixed transfer dispersion relation is that it breaks crossing symmetry. CSDR provides a dispersive representation which retains manifest crossing-symmetry. We consider elastic $2-2$ scattering of identical massive bosons of mass $m$ each in $d$ spacetime dimensions. The crossing symmetry of the corresponding scattering amplitude $\mm(s,t,u)$ translates into following invariance relation
\be 
\mm(s,t,u)=\mm(\p(s,t,u)),\qquad \forall\,\,\,\p\in S_3,
\ee $S_3$ being the group of permutations of three objects. 
 The basic argument is to write parametric dispersion relation on $S_3$ invariant cubic hypersurfaces in the space of complex Mandelstam variables $(s,t,u)$ subject to the on-shell constraint $s+t+u=4m^2$. For convenience, let us introduce the shifted Mandelstam variables $s_1=s-\m/3,\,s_2=t-\m/3,\,s_3=u-\m/3$. We consider the  family of cubic hypersurfaces labeled by a complex parameter $a$  
\be 
(a-s_1)(a-s_2)(a-s_3)=-a^3.
\ee This cubic can be rationally parametrized  as following 
\be \label{ratparam}
s_k(z,a):= a\left[1-\frac{(z-z_k)^3}{z^3-1}\right],
\ee $\{z_k\}$ being the cubic roots of unity. $z,\,a$ are crossing symmetric variables related to the shifted Mandelstam variables $(s_1,\,s_2,\,s_3)$ as $x=-(s_1s_2+s_2s_3+s_3s_1)=-27\,a^2\,z^3/(z^3-1)^2,\, y=-s_1s_2s_3=-27\,a^3\,z^3/(z^3-1)^2, \, a=y/x$. This parametrization maps the physical cuts $s_k\ge 2\m/3$ to the segments of unit circle in the complex $z$ plane. The parametrization \eqref{ratparam} enables to write a manifestly crossing symmetric dispersion relation as following 
\be \label{AKDR}
\mm( s_1, s_2)=\a_0+\int_{\frac{2\m}{3}}^\infty \, \frac{d\s}{\s}\, \ma(\s;\t^+(\s,a))\,H(\s;\, s_1, s_2,s_3),
\ee where $\ma(s_1,s_2)$ is the $s-$channel absorptive part of the amplitude, the discontinuity of the amplitude across the cut $s_1\ge\,\frac{2\m}{3}$  
\be 
\ma(s_1,s_2)=\lim_{\ve\to 0^+}\frac{1}{2i}\left[\mm(s_1+i\ve,s_2)-\mm(s_1-i\ve,s_2)\right],
\ee 
$\a_0=\mm(z=0, a)$ is the subtraction constant, and  
\begin{align}
\label{tpdef}	\t^+(\s,a)&:=-\frac{\s}{2}\left[1-\left(\frac{\s+3a}{\s-a}\right)^{\frac{1}{2}}\right],\\
\label{Hdef} H(\s;\, s_1,s_2,s_3)&:=\frac{s_1}{\s-s_1}+\frac{s_2}{\s-s_2}+\frac{s_3}{\s-s_3}
\end{align} 
\section{Dyson and Feynman blocks}
The crossing symmetric dispersion relation \eqref{AKDR}, along with partial wave expansion,  enable us to expand the scattering amplitude in terms of crossing-symmetric functions, the Dyson and the Feynman blocks. These were first introduced in \cite{ASAZ}. First we discuss them following  
\subsection{The Dyson block}
First we consider the partial wave expansion of the absorptive part $\ma(s_1,s_2):$
\be \label{partA}
\ma(s_1,s_2)=\Phi(s_1;\a)\sum_{\ell~\text{even}} (2\ell+2\a)\, a_\ell(s_1)\, C_\ell^{(\a)}\left(1+\frac{2s_2+2\m/3}{s_1-2\m/3}\right),
\ee where $C_\ell^{(\a)}(x)$ is the usual Gegenbauer polynomial, $\a=\frac{d-3}{2},\, \Phi(s_1;\a)=\Psi(\a)\, \frac{({s_1+\m/3})^{1/2}}{(s_1-2\m/3)^\a}$, $\Psi(\a)$ is a real positive number. We assume that the partial wave-expansion \emph{converges}. The details of the domain $\md$ in which the expansion converges will be described later. For now it suffices to assume the convergence. Next we insert the partial wave expansion \eqref{partA} into the crossing symmetric dispersion relation \eqref{AKDR} to write 
\be
\begin{aligned}
	&\mathcal{M}\left(s_{1}, s_{2}\right)=\alpha_{0}+\frac{1}{\pi} \int_{\frac{2 \mu}{3}}^{\infty} \frac{d \sigma}{\sigma}  H\left(\sigma ; s_{1}, s_{2}, s_{3}\right) \, \Phi(\sigma ; \alpha)\left[\sum_{\ell=0}^{\infty}(2 \ell+2 \alpha) \frac{a_{\ell}(\sigma)}{\left(\sigma-\frac{2 \mu}{3}\right)^{\ell}} Q_{\ell}\left(\sigma , \t^{(+)}(\sigma, a)\right)\right],
\end{aligned}
\ee
where 
\be
Q_{\ell}\left(s_{1}, s_{2}\right)=\left(s_{1}-\frac{2 \mu}{3}\right)^{\ell} C_{\ell}^{(\alpha)}\left(1+2 \frac{s_{2}+\frac{\mu}{3}}{s_{1}-\frac{2 \mu}{3}}\right).\ee Next using Fubini-Tonnelli theorem, we can exchange the order of integration over $\s$ and the infinite sum over $\ell$ to write the \emph{Dyson block expansion} of $\mm(s_1, s_2)$   
\be
\begin{aligned}
	&\mathcal{M}\left(s_{1}, s_{2}\right)=\alpha_{0}+\frac{1}{\pi} \sum_{\ell=0}^{\infty}(2 \ell+2 \alpha)\, \int_{\frac{2 \mu}{3}}^{\infty}\, d \sigma\,  \frac{\Phi(\sigma ; \alpha)}{\s\, \left(\sigma-\frac{2 \mu}{3}\right)^{\ell}} \,  a_{\ell}(\sigma)\, M_\ell^{(D)}(\s;s_1, s_2),
\end{aligned}
\ee where $M_\ell^{(D)}(\s;s_1, s_2)$ is the \emph{Dyson block} $M_\ell^{(D)}(\s,\t\,;\,s_1,s_2)$ defined by  
\be 
M_\ell^{(D)}(\s;s_1, s_2):= H(\s;s_1,s_2,s_3)\, Q_\ell(\s,\t^+(\s,a)).
\ee 
The manifest crossing-symmetry of the Dyson block comes at the cost of locality, i.e. the Dyson block contains \emph{non-local} terms with \emph{negative power} of the crossing-symmetric variable $x=-(s_1s_2+s_2s_3+s_3s_1)$. Thus if we want to work with Dyson block in local quantum field theories, we need to impose the \emph{locality constraints} \cite{ASAZ} to remove the negative powers of $x$. These locality constraints are quite involved, and are not easy to impose. A natural way to circumvent this would be to work in a basis of functions with the non-local negative powers removed at the onset. This leads to the definition of the \emph{Feynman block}.      

\subsection{The Feynman block and the contact terms}

\emph{Feynman block} is the Dyson block sans the  non-local terms, i.e. the terms with negative powers of $x$. The Feynman block $M_\ell^{(F)}(\s;s_1, s_2)$ can be expressed as 
 \be\label{eq:Feynmanstu}
M_{\ell}^{(F)}\left(\sigma ; s_{1}, s_{2}\right):=\sum_{i=1}^{3} M_{\ell}^{(i)}\left(\sigma ; s_{1}, s_{2}\right)+M_{\ell}^{(c)}\left(\sigma ; s_{1}, s_{2}\right)
\ee where
\be
\begin{split}
	&M_{\ell}^{(1)}\left(\sigma ; s_{1}, s_{2}\right)=Q_{\ell}\left(s_{1}, s_{2}\right)\left(\frac{1}{\sigma-s_{1}}-\frac{1}{\sigma}\right) \,,\\
	&M_{\ell}^{(2)}\left(\sigma ; s_{1}, s_{2}\right)= M_{\ell}^{(1)}\left(\sigma ; s_{2}, s_{3}\right), \quad M_{\ell}^{(3)}\left(\sigma ; s_{1}, s_{2}\right) = M_{\ell}^{(1)}\left(\sigma ; s_{3}, s_{1}\right),
\end{split}
\ee
and $M_\ell^{(c)}(\s;s_1,s_2)$ are the  \emph{contact terms} containing terms  \emph{polynomial} in $\{s_i\}$. The precise relation between Feynman and Dyson block is given by 
\be \label{DyFeyrel}
\frac{1}{\s}\,M_\ell^D(\s;s_1,s_2)-\sum_{i=1}^{3} M_{\ell}^{(i)}\left(\sigma ; s_{1}, s_{2}\right)=\mathcal{D}_{\ell}(\sigma;x,a),
\ee where $\mathcal{D}_{\ell}(\sigma;x,a)$ contains the non-local terms with negative powers of $x$. Structurally, $\mathcal{D}_{\ell}(\sigma;x,a)$ contains the non-local terms as well as the contact terms $M_\ell^{(c)}$. The non trivial terms in the Feynman block are the contact terms. We can   extract $M_\ell^{(c)}$ from   $\mathcal{D}_{\ell}(\sigma;x,a)$ by removing the non local terms systematically.
A general algorithm can be chalked out for this purpose as follows.  

{ First we expand the Gegenbauer polynomials as following:  
\be
Q_{\ell}\left(s_{1}, s_{2}\right)=\left(s_{1}-\frac{2 \mu}{3}\right)^{\ell} C_{\ell}^{(\alpha)}\left(1+2 \frac{s_{2}+\frac{\mu}{3}}{s_{1}-\frac{2 \mu}{3}}\right)=\sum_{p=0,q=0}^{\ell}\frac{b_{p,q}^{(\ell)}}{2}\left( s_1^p s_2^q+ s_1^p (-s_1-s_2)^q\right),
\ee where

\be
b_{p,r}^{(\ell)}=\frac{1}{p!\,r!}\partial_{s_1}^p\partial_{s_2}^r\,Q_\ell(0,0)=\sum_{k=0}^{\ell/2}\frac{(-1)^k 2^{-2 k-r+2 \ell } 3^{p+r-\ell } \binom{\ell -2 k}{p} (\alpha )_{\ell -k} (-\mu )^{-p-r+\ell } \binom{2 k}{2 k+p+r-\ell }}{k! (\ell -2 k)!}.
\ee
Using this expansion, we can write 
\be
\begin{split}
	 M_\ell^{(D)}(\s;s_1, s_2)&=\s\sum_{p=0,q=0}^{\ell}\frac{b_{p,q}}{2}I_{p,q}^{(D)}(\sigma,x,a)\,,\\
	\sum_{i=1}^{3} M_{\ell}^{(i)}\left(\sigma ; s_{1}, s_{2}\right)&=\sum_{p=0,q=0}^{\ell}\frac{b_{p,q}}{2}I_{p,q}^{(F)}(\sigma,x,a),
\end{split}
\ee
where
\begin{align}
I_{p,q}^{(D)}(\sigma,x,a)&=\frac{2^{-q} (-1)^q x (2 \sigma-3a )  \sigma ^{p+q-1}}{a x+\sigma ^3-\sigma  x}\left(\left(1-\sqrt{\frac{3 a+\sigma }{\sigma -a}}\right)^q+ \left(1+\sqrt{\frac{3 a+\sigma }{\sigma-a }}\right)^q\right),\\
	I_{p,q}^{(F)}(\sigma,x,a)&=\frac{3^{p+q+1} \left(\frac{a z}{z^3-1}\right)^{p+q+2}}{a \sigma  z \left(a x+\sigma ^3-\sigma  x\right)}\times\Big[\omega ^q \left(9 a^2 z^2+\sigma  (z-1)^2 \left(3 a z+\sigma  \left(z^2+z+1\right)\right)\right)\nn\\
	&\times (z-1)^p \left((z-\omega )^q+(\omega  z-1)^q\right)+ \frac{\omega ^p \left(3 a z-\sigma  \left(z^2+z+1\right)\right)}{z^2+z+1}\nn\\
	&\times \Big\{(\omega  z-1)^p \left(\omega ^q (z-\omega )^q+(z-1)^q\right) \left(3 a z \left(z^2+z+1\right)+\sigma  \omega  \left(z^3-1\right) (1-\omega  z)\right)\nn\\
	&\left.+(z-\omega )^p \left(\omega ^q (\omega  z-1)^q+(z-1)^q\right) \left(3 a z \left(z^2+z+1\right)-\sigma  \omega  \left(z^3-1\right) (z-\omega )\right)\right) \Big\}
	\Big].
\end{align}
Here $\omega=(-1)^{2/3}=-\frac{1}{2}+\frac{i \sqrt{3}}{2},\,z^3=\frac{2 x-3 a \left(\pm \sqrt{81 a^2-12 x}+9 a\right)}{2 x}$. The $I_{p,q}^{(F)}(\sigma,x,a)$ turns out to be a  rational function of $x, a$ for $p,q\in \mathbb{Z}^{\geq }$. We leave the detailed proof for this fact to the upcoming sections. For now, we can see this in a few explicit examples as following: 
\be
\begin{split}
	&I_{0,1}^{(F)}(\sigma,x,a)=\frac{x (3 a-2 \sigma )}{a x+\sigma ^3-\sigma  x}\,, I_{1,0}^{(F)}(\sigma,x,a)=\frac{4 \sigma  x-6 a x}{a x+\sigma ^3-\sigma  x}\,, I_{1,1}^{(F)}(\sigma,x,a)=\frac{x \left(3 a \sigma ^2+2 a x-2 \sigma  x\right)}{\sigma  \left(a x+\sigma ^3-\sigma  x\right)}\,,\\
	&I_{10,0}^{(F)}(\sigma,x,a)=\frac{2 x^4 \left(a^3 \sigma  (3 a-11 \sigma )+a x \left(-15 a^2+24 a \sigma -11 \sigma ^2\right)+2 x^2 (\sigma -a)\right)}{\sigma  \left(a x+\sigma ^3-\sigma  x\right)}.
\end{split}
\ee
Collecting everything together, we can write 
\be\label{DelltoIDF}
\mathcal{D}_{\ell}(\sigma;x,a)=\sum_{p=0,q=0}^{\ell}\frac{b_{p,q}}{2}\left(I_{p,q}^{(D)}(\sigma,x,a)-I_{p,q}^{(F)}(\sigma,x,a)\right )
\ee
While at the outset $\md_\ell$ looks pretty complicated, it can be written in a suggestive form as
\be\label{Dell_obser}
\mathcal{D}_{\ell}(\sigma;x,a)=\frac{1}{(\s-a)^{\ell/2}}\times \mathfrak{P} (x,a)
\ee
where $\mathfrak{P}(x,a)$ is a polynomial of the form\footnote{It is very hard to find a general formula for $\mathfrak{P} (x,a)$, we work with case by case in $\ell$. We will not need the general formula.} 
\be 
\mathfrak{P}(x,a)=\sum_{M=0}^{M_1}\sum_{N=0}^{N_1}\, \mathfrak{P}_{MN}\, x^M\, a^N 
\ee with $M_1\le \ell/2,\,  N_1\le\frac{1}{9} \left(\frac{15 \ell }{2}-\sqrt{3} \sin \left(\frac{\pi  \ell }{3}\right)+3 \cos \left(\frac{\pi  \ell }{3}\right)-3\right)$. Next substituting $a=y/x$ into ${(\s-a)^{-\ell/2}}$ we expand the same to obtain a Laurent series about $x=0$.  The terms $x^{-\tilde{M}}, \, \tilde{M}>\ell/2$ in the Laurent series will contribute towards the  non-local terms. We can throw away these terms to obtain the contact terms $M_\ell^c$. 

 We work out here $\ell=2$ case as an example. Consider $\mathcal{D}_{\ell=2}(\sigma;x,a)=\frac{x b_{0,2}^{(2)}}{a-\sigma}+\frac{x\left(b_{0,2}^{(2)}+2 b_{2,0}^{(2)}\right)}{\sigma}=$ $\frac{2 x b_{2,0}^{(2)}}{\sigma}-\frac{y b_{0,2}^{(2)}}{(\sigma)^{2}}-\left(x b_{0,2}^{(2)}\right) \sum_{n=2}^{\infty} a^{n}(\sigma)^{-n-1}$. Now throwing away negative powers in $x$ (non-local terms), we get
$$
M_{\ell=2}^{(c)}\left(\sigma ; s_{1}, s_{2}\right)=\frac{2 x b_{2,0}^{(2)}}{\sigma}-\frac{y b_{0,2}^{(2)}}{\sigma^{2}}.
$$
This is the \emph{expected} polynomial form for the contact terms. Similarly, we can work for other $\ell$. 

The suggestive form \eqref{Dell_obser} enables us to compute the contact terms in complete generality. Since $\mathcal{D}_\ell(\sigma;x,a)$ is a polynomial in $x$ upto order $\ell/2$,   we can only keep the terms upto $x^{\ell/2}$ in \eqref{DelltoIDF}. We would like to stress that $\left(I_{p,q}^{(D)}(\sigma,x,a)-I_{p,q}^{(F)}(\sigma,x,a)\right )$ will have all kinds of powers of $x$, since there is a pole at  $\left(a x+\sigma ^3-\sigma  x\right)=0$. After truncating the power of $x$ upto $x^{\ell/2}$, we will then have to remove the non-local terms from the contribution of $I_{p,q}^{(D)}(\sigma,x,a)$. One can see that contributions from $I_{p,q}^{(F)}(\sigma,x,a)$ is always local. Consider $n^{th}$ partial sum for the Taylor series of $I_{p,q}^{(F/D)}(x,a)$
\be
\mathcal{J}_{n,p,q}^{(F/D)}(x,a):=\sum_{j=0}^{n}\frac{x^j}{j!} ~~\left[\frac{\partial^j}{\partial x^j}\,I_{p,q}^{(F/D)}(0,a)\right].
\ee

Next denote by $\mathcal{J}_{n,m,p,q}^{(D)}(\sigma,x,a)$ the $x^n a^m$ term in the double series expansion of   $I_{p,q}^{(D)}(\sigma,x,a)$ in $(a,x)$. We only keep local terms in $a$ which implies $m\leq n$. 

Collecting everything together,  the contact terms can be generically expressed as  
\be\label{eq:Feynmancm}
\boxed{M_{\ell}^{(c)}\left(\sigma ; s_{1}, s_{2}\right)=\sum _{p=0}^{\ell } \sum _{q=0}^{\ell } \frac{b_{p,q}^{(\ell)}}{2}\left[-\mathcal{J}_{\frac{\ell }{2},p,q}^{(F)}(\sigma,x,a)+\sum _{n=1}^{\frac{\ell }{2}} \sum _{m=0}^n \mathcal{J}_{n,m,p,q}^{(D)}(\sigma,x,a)\right]\,.}
\ee

\subsubsection{Derivation of $\mathcal{J}_{n,p,q}^{(F)}(\sigma,x,a)$ }\label{derivationJF}
We first explore the from of $I^{(F)}_{p,q}(x,a)$
\be
I^{(F)}_{p,q}(x,a)=f_{p,q}(z)+f_{p,q}(\omega z)+f_{p,q}(\omega^2 z)
\ee
Note that $s$-channel Feynman block is related to  $f_{p,q}(z)$ via
\be
M_{\ell}^{(1)}\left(\sigma ; s_{1}, s_{2}\right)=Q_{\ell}\left(s_{1}, s_{2}\right)\left(\frac{1}{\sigma-s_{1}}-\frac{1}{\sigma}\right)=\sum_{p=0,q=0}^{\ell}\frac{b_{p,q}}{2}f_{p,q}(z)
\ee
Similarly
\be
M_{\ell}^{(2)}\left(\sigma ; s_{1}, s_{2}\right)=\sum_{p=0,q=0}^{\ell}\frac{b_{p,q}}{2}f_{p,q}(\omega^2 z)\,,~M_{\ell}^{(3)}\left(\sigma ; s_{1}, s_{2}\right)=\sum_{p=0,q=0}^{\ell}\frac{b_{p,q}}{2}f_{p,q}(\omega z)
\ee

The form of $f_{p,q}(z)$ can be easily found to be
\be
\begin{split}
f_{p,q}(z)=&\frac{a z 3^{m+n+1}  \left(\frac{a z}{z^3-1}\right)^{m+n}(z-1)^n}{\sigma  \left(-27 a^3 z^3+27 a^2 \sigma  z^3+\sigma ^3 \left(z^3-1\right)^2\right)}\times\\
&\left(9 a^2 z^2+\sigma  (z-1)^2 \left(3 a z+\sigma  \left(z^2+z+1\right)\right)\right) \left(\omega ^m \left((z-\omega )^m+(\omega  z-1)^m\right)\right)
\end{split}
\ee

We can write the above one in power series in the following form
\be
f_{p,q}(z)=\frac{a 3^{p+q+1} z^{p+q+1} \left(\frac{a}{z^3-1}\right)^{p+q} \sum _{h=0}^4 \sum _{r=0}^{p+q} \Lambda (h) z^{h+r} \mathcal{S}_{r,p,q}}{\sigma  \left(-27 a^3 z^3+27 a^2 \sigma  z^3+\sigma ^3 \left(z^3-1\right)^2\right)},
\ee
where 
\be\label{SS}
\mathcal{S}({r,p,q})=(-1)^{r+p+q} \binom{q}{r} \left[ \omega ^{-r-q} \, _2F_1(-p,-r;q-r+1;\omega )+\left(\omega^2 \right)^{-r-q} \, _2F_1\left(-p,-r;q-r+1;\omega ^2\right)\right].
\ee
One has that $\mathcal{S}_{r,p,q}\in \mathbb{R}$. $\Lambda(h)$ is the coefficient $z^h$ in the expression $9 a^2 z^2 + 
3 a (-1 + z)^2 z \s + (-1 + z)^2 (1 + z + z^2) \s^2$, which is given by
\be\label{Phi}
\Lambda(h)= 
\begin{cases}
	3 a (3 a-2 \sigma ) & h=2 \\
	\sigma  (3 a-\sigma ) & h=1,\, h=3 \\
	\sigma ^2 & h=0,\, h=4 \\
\end{cases}
\ee

Next we will use a simple result concerning the cube roots of unity  $\{1, \, \omega, \, \omega^2\}$. Consider a convergent power series
\be
f(x)=\sum_{r=r_{\min}}^{r_{\max} }a_{r} x^{r}.
\ee
Then we have 
\be\label{cubertthm}
\frac{f(x)+f\left(\omega x\right)+f\left(\omega^2 x\right)}{3}=\sum_{r=\lfloor \frac{r_{\min}}{3 }\rfloor}^{\lfloor \frac{r_{\max}}{3 }\rfloor } a_{3 r} x^{3 r}
\ee To prove this, we can always write  the monomial  $x^k$ as $x^{3r+p}$, where $r\in\mathbb{Z},\, p\in\{0,\,1,\,2\}.$ Using this we have 
\be \label{monomialthm}
\frac{1}{3}\left[x^k+(\omega\,x)^k+(\omega^{2}\, x)^{k}\right]=\frac{x^{3r+p}}{3}\left[1+\omega^p+\omega^{2p}\right]=x^{3r+p}\, \delta_{p,0},
\ee where in the last step we have used the properties of the cube roots of unity
$$
\frac{\left(1^{n}+\omega^{n}+\omega^{2n}\right)}{3}= \begin{cases}1, & \text { if }\, 3|n \\ 0, & \text { else }\end{cases}.
$$ \eqref{cubertthm} then follows immediately from \eqref{monomialthm}.

Using this result, we can write
\be
I^{(F)}_{p,q}(x,a)=3\frac{a 3^{p+q+1} \left(\frac{a}{z^3-1}\right)^{p+q} \sum _{h=0}^4 \sum _{r=\left\lfloor \frac{1}{3} (h+p+q+1)\right\rfloor }^{\left\lfloor \frac{1}{3} (h+2p+2q+1)\right\rfloor} \Lambda (h) z^{3 r} \mathcal{S}_{3r-h-p-q-1,p,q}}{\sigma  \left(-27 a^3 z^3+27 a^2 \sigma  z^3+\sigma ^3 \left(z^3-1\right)^2\right)},
\ee
which can be further expressed as 
\be
\begin{split}
I^{(F)}_{p,q}(x,a)=&3\times \sum_{i=0}^{p+q-2}\left(\frac{a \left(1-z^3\right)^2 3^{p+q+1} (-1)^{i+p+q} \left(\frac{a z^3}{\left(z^3-1\right)^2}\right)^{p+q} \binom{p+q-2}{i} z^{3 i-3 (p+q)}}{\sigma  \left(-27 a^3 z^3+27 a^2 \sigma  z^3+\sigma ^3 \left(z^3-1\right)^2\right)}\right)\\
& \times \left ( \sum _{h=0}^4 \sum _{r=\left\lfloor \frac{1}{3} (h+p+q+1)\right\rfloor }^{\left\lfloor \frac{1}{3} (h+2p+2q+1)\right\rfloor} \Lambda (h) z^{3 r} \mathcal{S}_{3r-h-p-q-1,p,q}\right).
\end{split}
\ee
The above formula is not valid for $p+q\leq 1$. 
Now we note that
\be\label{z3j}
z^{-3j}=27^j \left(-\frac{x}{a^2}\right)^{-j} \, _2F_1\left(\frac{1}{2}-j,-j;1-2 j;\frac{4 x}{27 a^2}\right)
\ee
This can be seen from the formula 
$$
z^{3}=\nu\left[\frac{1-(1+4 \nu)^{1 / 2}}{2 \nu}\right]^{2} \,, \nu=\frac{z^3}{\left(z^3-1\right)^2}
$$
We have to do the expansion of the $2 j$th power of the term in the bracket above. This is  related  the so-called Catalan numbers
$$
\left[\frac{1-(1+4 \nu)^{1 / 2}}{2 \nu}\right]^{2 n}=\sum_{m=0}^{\infty} \frac{n}{(m+n)}\left(\begin{array}{c}
2 m+2 n \\
m
\end{array}\right)(-1)^{m} \nu^{m} .
$$

\be
z^{3 j}=\sum _{i=0}^{\infty } \frac{(-1)^i j \left(\frac{z^3}{\left(z^3-1\right)^2}\right)^i \binom{2 i+2 j}{i}}{i+j}
\ee
and we use the fact that our $I^{(F)}_{p,q}(x,a)$ has a symmetry under exchange of $z\to 1/z$ \eqref{z3j}. 
Therefore we can write
\be\label{IFseries}
\begin{split}
&I^{(F)}_{p,q}(x,a)=\sum _{h=0}^4 {\sum}_{j=0}^{\infty}{\sum} _{r=\left\lfloor \frac{h+p+q+1}{3} \right\rfloor }^{\left\lfloor \frac{h+2 p+2 q+1}{3}\right\rfloor }{\sum} _{i=0}^{p+q-2} {\sum} _{k=0}^{p+q-r-i}\Bigg[\frac{4^k (-1)^{p+q+r} a^{p+q+1-2 i-2 k-2 r}}{3^{2-p-q}\sigma ^4 }\left(\frac{27 (\sigma -a)}{\sigma ^3}\right)^j\\
&\times  \left(\frac{x}{27}\right)^{i+j+k+r}\Lambda (h)  \mathcal{S}(-h-p-q+3 r-1,p,q)\frac{ (2-p-q)_i (i+r-p-q)_k \left(i+r-p-q+\frac{1}{2}\right)_k}{k! i! (2 i-2 p-2 q+2 r+1)_k}\Bigg]\\
\end{split}
\ee
The $j$-sum comes for expanding $1/\left(a x+\sigma ^3-\sigma  x\right)$ and $k$-sum comes from expanding the ${}_2F_1$. The above formula is not valid for $p+q\leq 1$. We have to work out the power till $x^n$ by hand for $p=0,q=0,~ p=1,q=0,~ p=0,p=1$. Lets denote the expansion of $I_{p,q}^{(F)}(\sigma,x,a)$ upto $x^{n}$ as $\mathcal{J}_{n,p,q}^{(F)}(\sigma,x,a)$. This gives the final formula 
\be\label{JFM}
\begin{split}
&\mathcal{J}_{n,p,q}^{(F)}(\sigma,x,a)=\sum _{h=0}^4 \widehat{\sum}_{j=0}^{n}\widehat{\sum} _{r=\left\lfloor \frac{h+p+q+1}{3} \right\rfloor }^{\left\lfloor \frac{h+2 p+2 q+1}{3}\right\rfloor }\widehat{\sum} _{i=0}^{p+q-2} \widehat{\sum} _{k=0}^{p+q-r-i}\Bigg[\frac{4^k (-1)^{p+q+r} a^{p+q+1-2 i-2 k-2 r}}{3^{2-p-q}\sigma ^4 }\left(\frac{27 (\sigma -a)}{\sigma ^3}\right)^j\\
&\times  \left(\frac{x}{27}\right)^{i+j+k+r}\Lambda (h)  \mathcal{S}(-h-p-q+3 r-1,p,q)\frac{ (2-p-q)_i (i+r-p-q)_k \left(i+r-p-q+\frac{1}{2}\right)_k}{k! i! (2 i-2 p-2 q+2 r+1)_k}\Bigg]\\
&+~\sum _{j=1}^n \frac{(3 a-2 \sigma ) \left(\frac{\sigma -a}{\sigma ^3}\right)^j (2 \delta _{0,q} (\sigma  \delta _{1,p}+\delta _{0,p})-\sigma  \delta _{0,p} \delta _{1,q})}{\sigma  (a-\sigma )}
\end{split}
\ee
The wide hat in the summation of $i,j,k,r$ is to denote that the sums have to be restricted such that $i+j+k+r\leq n$.

\subsubsection{Derivation of $\mathcal{J}_{n,m,p,q}^{(D)}(\sigma,x,a)$ }\label{derivationJD}
We can write $I_{p,q}^{(D)}(\sigma,x,a)$ in terms of Chebyshev $T_n(x)$
\be
\begin{split}
I_{p,q}^{(D)}(\sigma,x,a)=&\frac{2^{-q} (-1)^q x (2 \sigma-3a )  \sigma ^{p+q-1}}{a x+\sigma ^3-\sigma  x}\left(\left(1-\sqrt{\frac{3 a+\sigma }{\sigma -a}}\right)^q+ \left(1+\sqrt{\frac{3 a+\sigma }{\sigma-a }}\right)^q\right)\\
&=\frac{2^{-q} x (2 \sigma-3a )  \sigma ^{p+q-1}}{a x+\sigma ^3-\sigma  x}\ 2^{q+1} \left(\frac{\sqrt{a-\sigma }}{\sqrt{a}}\right)^{-q} T_q\left(\frac{\sqrt{a-\sigma }}{2 \sqrt{a}}\right)
\end{split}
\ee
We can expand the $T_q\left(\frac{\sqrt{a-\sigma }}{2 \sqrt{a}}\right)$ in powers of $a$, which gives
\be
I_{p,q}^{(D)}(\sigma,x,a)=\frac{2^{-q} x (2 \sigma-3a )  \sigma ^{p+q-1}}{a x+\sigma ^3-\sigma  x}\  \sum_{r=0}^{\infty} \sum_{k=0}^{q/2}2 (-1)^{q-r}  \left(\frac{a}{\sigma }\right)^r \binom{q}{2 k} \binom{-k}{r} \, _2F_1(-k,-r;-k-r+1;-3)
\ee
Now we can expand the above in powers of $x$, which gives
\be
I_{p,q}^{(D)}(\sigma,x,a)=\sum_{n=0}^{\infty}\left(\frac{ (2 \sigma -3 a) (\sigma -a)^{n-1}}{\sigma ^{3 n+1-p-q}2^q} \right) \sum_{r=0}^{\infty} \sum_{k=0}^{q/2}2 (-1)^{q-r}  \left(\frac{a}{\sigma }\right)^r \binom{q}{2 k} \binom{-k}{r} \, _2F_1(-k,-r;-k-r+1;-3)
\ee

Now we can again expand in powers of $a$, after rearranging the sum a little, we get 

\be
I_{p,q}^{(D)}(\sigma,x,a)=\sum_{n=0}^{\infty}\sum_{m=0}^{\infty} \sum_{r=0}^{m}\sum_{k=0}^{\lfloor q/2 \rfloor}\frac{2^{1-q} (-1)^{q+m} \Gamma (n) \binom{q}{2 k} \binom{-k}{r} \, _2F_1(-k,-r;-k-r+1;-3) (m+2 n-r) }{\sigma ^{2 n+m+1-p-q}\Gamma (m-r+1) \Gamma (-m+n+r+1)}a^m x^n
\ee
Lets denote by $\mathcal{J}_{n,m,p,q}^{(D)}(\sigma,x,a)$ the $x^n a^m$ term in the double series expansion of   $I_{p,q}^{(D)}(\sigma,x,a)$ in $(a,x)$. This basically gives 
\be\label{JDM}
\mathcal{J}_{n,m,p,q}^{(D)}(\sigma,x,a)=\sum_{r=0}^{m}\sum_{k=0}^{\lfloor q/2 \rfloor}\frac{2^{1-q} (-1)^{q+m} \Gamma (n) \binom{q}{2 k} \binom{-k}{r} \, _2F_1(-k,-r;-k-r+1;-3) (m+2 n-r) }{\sigma ^{2 n+m+1-p-q}\Gamma (m-r+1) \Gamma (-m+n+r+1)}a^m x^n
\ee
}

\section{Analyticity domains of scattering amplitudes: Axiomatic results}
In the previous section, we introduced the structure of Dyson and Feynman block expansions of the scattering amplitude. We will now take up the discussion of convergence properties. These convergence properties play central role in analyzing the analyticity properties of the scattering amplitude $\mm$.  In particular, we would like to analyze the convergence properties of the Feynman block expansions. First we give a compact review of key results concerning the convergence properties of the usual partial wave expansion of the scattering amplitude. We also provide some discussion on the analyticity domains of the scattering amplitude that have been derived in the framework of axiomatic quantum field theory.  

\subsection{Convergence of partial wave expansion}
We consider the $s-$channel partial wave expansion 
\begin{align}
	\mm(s_1,s_2)&=\Phi(s_1;\a)\sum_{\ell~\text{even}} (2\ell+2\a)\, f_\ell(s_1)\, C_\ell^{(\a)}\left(x\right),\\
	\ma(s_1,s_2)&=\Phi(s_1;\a)\sum_{\ell~\text{even}} (2\ell+2\a)\, a_\ell(s_1)\, C_\ell^{(\a)}\left(x\right);
\end{align}
with \be 
a_\ell(s):=\text{Im.}\,f_\ell(s)=\frac{1}{2i}\,\left[f_\ell(s_1+i\, 0^+,s_2)-f_\ell(s_1-i\, 0^+,s_2)\right],\qquad x:=1+\frac{2s_2+2\m/3}{s_1-2\m/3}.
\ee Here $C_\ell^{(\a)}$ are the Gegenbauer polynomials with $\a=(d-3)/2$, and $\Phi(s_1;\a)=\Psi(\a) \sqrt{s_1+\m/3}/(s_1-2\m/3)^\a$, $\Psi(\a)$ being a positive number depending only on $\a$.   While the dispersive representation guarantees analyticity in complex cut $s_1$ plane, we are interested in analyticity in complex $s_2$. Analyticity in a certain domain in complex $s_2$ plane will automatically imply convergence of the partial wave expansion in the corresponding domain of complex $x$ plane.  Let us give a brief review of key results that have been proved at various stages in the framework of axiomatic quantum field theory \emph{a-la} Lehmann, Bros, Epstein, Glaser, Martin etc. 

\paragraph{}  The very structure of partial wave expansion in terms of the Gegenbauer polynomial suggests certain analytic structure of $\mm(s_1, s_2)$ in complex $s_2$. An argument due to Neumann \cite{WhitaWat} shows that a Gegenbauer series, i.e. a series of the form $\sum_n \a_n\, C_n^{(\a)}(w)$, converges in some open elliptical disc  with the boundary ellipse having foci at $w=\pm 1$. The exact shape of the ellipse, i.e. the major and minor axes, depends upon the magnitude of coefficients $\{\a_n\}$.

Thus one can anticipate, for both the amplitude $\mm$ and the absorptive part $\ma$, elliptic domains of analyticity in complex $z-$plane, and, therefore, in complex $s_2-$plane for fixed $s_1$. Starting from the LSZ axiomatics, Lehmann \cite{Lehmann1} first showed the existence of such elliptical domains for \textit{physical} $s_1$.  It follows that, for physical  $s_1\ge 2\m/3$, the scattering amplitude $\mm(s_1,x)$ is analytic in complex $x-$plane inside  an ellipse with foci at $\pm 1$ and semi-major axis 
\be \label{LehmannS}
x_s=\left[1+\frac{4(M^2-m^2)^2}{s(s-\m)}\right]^{\frac{1}{2}},
\ee 
 $M=2m$ for identical scalar bosons and $M=3m$ for pions. This ellipse is called \textit{small Lehmann ellipse}.  Lehmann further showed that the absorptive part $\ma(s_1, x)$ is analytic  in a larger open elliptic disc bounded by the so-called \textit{large Lehmann ellipse}, $\me_L(s)$. The semi-major axis of this larger ellipse is related to that of the small Lehmann ellipse by 
\be \label{LehmannL}
x_l=2x_s^2-1=1+\frac{8(M^2-m^2)^2 }{s(s-\m)}.
\ee 
However these analyticity domains can be extended further using unitraity. This was largely achieved by Martin as we will discuss now. 

\subsection{Martin's extension of analyticity domain}
The key ingredient in Martin's \cite{MartinExt1} extension of analyticity domains for $\mm$ and $\ma$ is \emph{unitarity}. It is to be stressed that so far unitarity has not been used in deriving the  Lehmann ellipses.  Lehmann only used  analytic consequences of micro-causality.  Let us first state Martin's theorem on the extension of domain, and then we will discuss the important consequences of this theorem. 
\begin{theorem}[{\bf Martin's extension theorem}]\label{MartinExt}
	Consider an \textbf{unitary}  elastic scattering amplitude $\mm(s_1,s_2)$, and the corresponding absorptive part $\ma(s_1,s_2)$, for the process $A+A\to A+A$ of massive particle $A$ with mass $m>0$ satisfying analyticity properties:
	\begin{itemize}
		\item [I.] {\bf Dispersive representation: }$\mm(s_1,s_2)$ satisfies fixed $t$ dispertion relations for $-s_{2}^{(0)}<s_2\le0,\,s_2^{(0)}>0$ where the amplitude is \textbf{analytic in complex $s_1$ plane} outside of the unitarity cuts;
		\item[II.] {\bf Lehmann analyticity: }For \textbf{fixed physical} $s_1$, the amplitude $\mm(s_1,s_2)$ and the absorptive part $\ma(s_1,s_2)$ are analytic in the corresponding Lehmann ellipses;
		\item[III.] {\bf Bros-Epstein-Lehmann-Glaser  (BELG) analyticity:} From the results of Bros, Epstein and Glaser, and Lehmann \cite{BELG}, in the neighbourhood of any point $s_1', s_2'$, $-s_2^{(0)}<s_2'\le0$, $s_1'$ outside the cuts, there is analyticity in both $s$ and $t$ in 
		\be 
		|s-s_0|<\eta(s_0,t_0),\quad |t-t_0|<\eta(s_0,t_0).
		\ee  
		A priori the size of the neighbourhood can vary with $s_0,t_0$.
	\end{itemize} 
Then the amplitude $\mm(s,t)$ is analytic (except for possible fixed-$t$ poles) in the topological product domain 
\be 
\mathcal{D}_M:=\{(s,t):s\not\in \left[\left(4m^2,\infty\right)\cup\left(-\infty,-t\right)\right]\,\times\, |t|<R\}, 
\ee $R$ being independent of $(s,t)$. The absorptive part $\ma_{s}(s,t)$ is also analytic in $|t|<R$ for all $s$. 
 
\end{theorem}

Unitarity plays key role in proving Martin's theorem. Unitarity is used in the form of positivity. In particular, due to unitarity one has 
\be 
a_\ell(s)\ge 0\quad  \forall\, \ell, s. 
\ee Further, using the properties of Legendre polynomial, unitarity results into 

\begin{align}
\frac{d^n}{dz^n}\,\ma_s(s,z=1)&\ge 0\\
\frac{d^n}{dz^n}\,\ma_s(s,z)&\le \frac{d^n}{dz^n}\,\ma_s(s,z=1),\qquad z\in[-1,1].
\end{align} 
These positivity properties enable one to prove Martin's theorem along with the provided analytic structures. It is also worth the mention that one can prove this theorem \emph{without} assuming the existence of fixed $t$ dispersion relation. See \cite{MartinLec} for the detailed argument.

The greatest significance of Martin's extension theorem is that it shows existence of analyticity domain in $t$ for both $\mm$ and $\ma_s$ which \emph{does not} vary with $s$. In particular, as evident from \eqref{LehmannS}, \eqref{LehmannL}, the Lehmann domains in $t-$plane shrink to zero as $|s|\to\infty$.  Martin's theorem establishes that even when $|s|\to \infty$ there exist finite, \emph{non-vanishing} domains of analyticity in $t$ for \emph{both} $\mm(s,t)$ and $\ma_s(s,t)$.  As a consequence of this theorem we can find out an enlarged domain in $t-$plane where the fixed-$t$ dispersion relations hold. For the rest of the discussion we will consider elastic scattering of pions. In this case $R=\m$. Then we have that $\ma_s(s,t)$ is analytic in the open disc $|t|<\m$. However, $\ma_s$ is also expendable in terms of Legendre polynomials with \emph{positive coefficients} (due to unitarity). Now a Legendre polynomial expansion with positive coefficients \emph{must} have a singularity at the extreme right of the largest ellipse of convergence. Hence $\ma_s(s,t)$ having no singularities for $0\le t<R$ must be analytic in an open elliptical disc bounded by an ellipse  with foci at $t=0, \,\m-s$ and right extremity at $\m$. We will call this ellipse  \emph{the unitarity ellipse}, $\me_U(s)$. But, $\ma_s(s,t)$ is also analytic in the large Lehmann ellipse $\me_L(s)$ with  foci at $t=0, \m-s$ and right extremity at $16\m^2s^{-1}$. 
Then the required enlarged domain is given by 
\be
\md=\bigcap_{\m\le s<\infty} \left(\me_L(s)\cup\me_U(s)\right).
\ee
From the dimensions of the ellipse, it is quite evident that $\me_U(s)\subset\me_L(s)$ for $s< 16\m$, and $\me_L(s)\subset\me_U(s)$ for $s> 16\m$. Also $\me_U(16\m)\subset\me_U(s)$ for $s\ge 16\m$, and $\me_L(4\m)\subset \me_L(s)$ for $s\in [\m,4\m]$. Thus finally
\be 
\md=\bigcap_{4\m\le s\le16\m}\me_L(s).
\ee

 
The domains of $s_2$ analyticity can be extended further using elastic unitarity relation 
\be 
|f_\ell(s_1)|=a_\ell(s_1),\qquad s_1\in\left[\frac{2\m}{3}, \frac{11\m}{3}\right].
\ee Using this Martin \cite{MartinExt2} proved the existence of following analyticity domain in complex $s_2$ plane 
 
The details of Lehmann-Martin ellipses for the various regions and plots can be found in our appendix \eqref{sec:Lehmann-Martin ellipses}. 

\section{Analyticity domain in complex $a$-plane}
\begin{figure}[hbt!]
\centering
\includegraphics[scale=0.5]{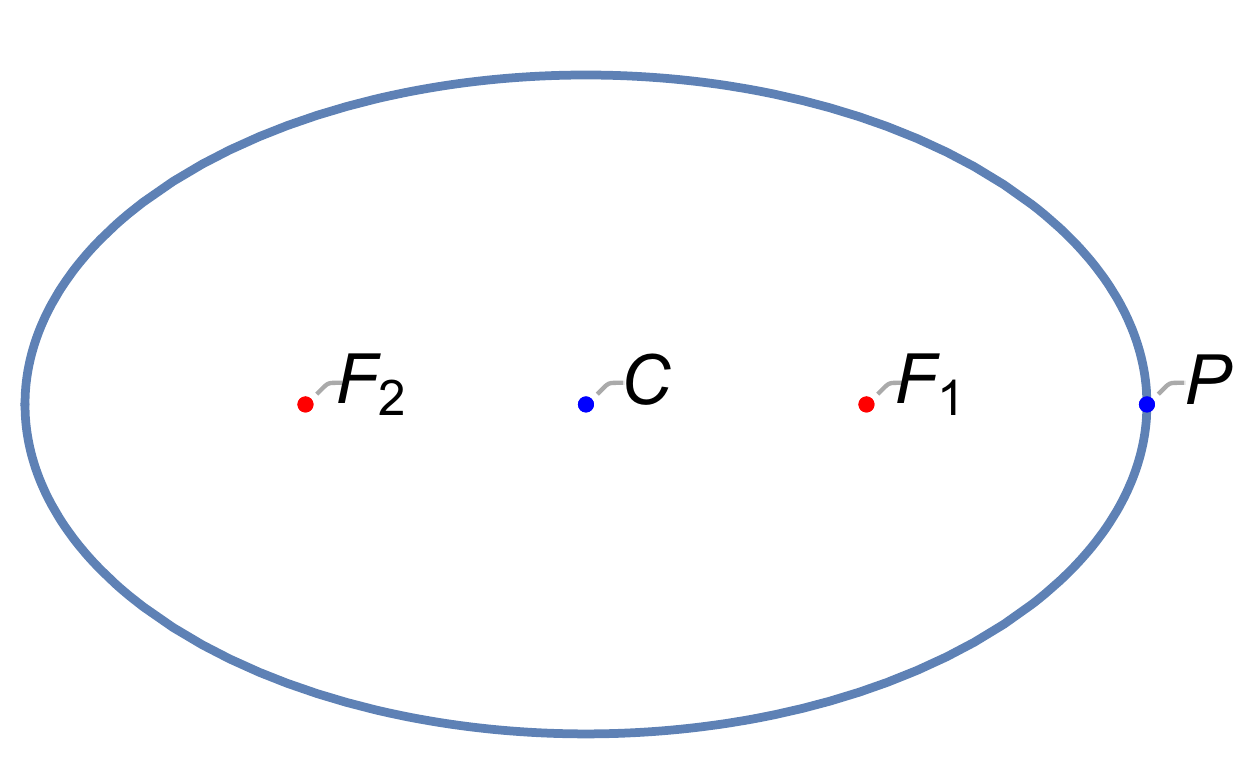}
\caption{An ellipse described by the equation $
\frac{(x-h)^2}{A^2}+\frac{(y-k)^2}{B^2}=1
$ with \emph{eccentricity} $e^{2}=1-\frac{B^2}{A^2}$. The foci $F_1$ and $F_2$ have the planar coordinates $(h+A e,k)$ and  
$\left(h-A e, k\right)$ respectively. Referred to this configuration, the right extremity $P$ is given coordinized by  
$(d=A+A e, k)$.}
\label{fig:genelip}
\end{figure}
We will now translate the analyticity domains in terms of the variable $a$. 
To derive the analyticity domain in complex $a$-plane, we will have to work with the Lehmann-Martin ellipses, and from those, we derive their analogous equation of the curves in complex $a$-plane. 
The analyticity domain for the absorptive part  in complex $s_2$ for various range of $s_1$, the so called \emph{Lehmann-Martin ellipse},  is given by  
\be
\label{eq:martin}
\me\left(s_{1}\right)=\begin{cases}
E\left(-\frac{\m}{3}, \frac{ \mu}{3}-s_{1} \mid \frac{11 \mu}{3}+\frac{12 \mu^2}{3 s_{1}-2 \mu}\right), & \frac{2 \mu}{3}<s_{1}<\frac{1 \mu}{3} \\
\\
E\left(-\frac{\m}{3}, \frac{ \mu}{3}-s_{1} \mid \frac{48 \mu^2}{3 s_{1}+\mu}-\frac{\m}{3}\right), & \frac{11 \mu}{3}<s_{1}<\frac{23 \mu}{3} \\
\\
E\left(-\frac{\m}{3}, \frac{ \mu}{3}-s_{1} \mid \frac{2\mu}{3}+\frac{12 \mu^2}{3 s_{1}-11 \mu}\right), & s_{1}>\frac{23 \mu}{3}
\end{cases}.
\ee
where $E\left(f_{1}, f_{2} \mid d\right)$ stands for an ellipse with foci at $s_{2}^{+}=f_{1}, s_{2}^{+}=f_{2}$ and right extremity at $s_{2}^{+}=d $. 
 We follow the convention that $f_1$ corresponds to the \emph{right} focus and $f_2$ corresponds to the \emph{left} focus \emph{with respect to} the center. 
%
Then, referred to the above figure, we have for $\me(s_1)$,
$F_{1}:(f_1,0)$, 
$F_{2}:\left(f_{2}, 0\right)$, 
$P:(d, 0)$, 
$C:\left(\frac{f_1+f_{2}}{2}, 0\right)$. Now from the simple geometric structure of the ellipse, we have 
\be
A e+A+f_2=d \text{ and } f_{1}=\frac{f_1+f_2}{2}+A e~~ \Rightarrow ~A=d-\frac{f_1+f_2}{2}, ~~e=\frac{-f_1+f_{2}}{\left(2 d-f_1-f_{2}\right)}
\ee
which gives the following equation for curve in complex $a$ plane
\be
\label{eq:masterdomain}
\frac{\left(\Re(s_2^+)-f_{2} / 2\right)^{2}}{\left(d-\frac{f_1+f_{2}}{2}\right)^{2}}+\frac{\Im(s_2^+)^{2}}{(d -f_1)(d - f_2)}=1
\ee
with
\be
\begin{split}
&\Re(s_2^+)=-\frac{s_1}{2}+\frac{s_1 \cos \left(\frac{1}{2} \arg \left(\frac{s_1+3 \Re(a)+3 i \Im(a))}{s_1-\Re(a)-i \Im(a)}\right)\right)}{2 \sqrt[4]{\frac{\Im(a)^2+\left(s_1-\Re(a)\right){}^2}{9 \Im(a)^2+\left(3 \Re(a)+s_1\right){}^2}}}\,, \Im(s_2^+)=\frac{s_1 \sin \left(\frac{1}{2} \arg \left(\frac{s_1+3 \Re(a)+3 i \Im(a))}{s_1-\Re(a)-i \Im(a)}\right)\right)}{2 \sqrt[4]{\frac{\Im(a)^2+\left(s_1-\Re(a)\right){}^2}{9 \Im(a)^2+\left(3 \Re(a)+s_1\right){}^2}}}
\end{split}
\ee
We have the analyticity domain in complex $a$ plane which governed by the above curve. The domain of analyticity is $\frac{\left(\Re(s_2^+)-f_{2} / 2\right)^{2}}{\left(d-\frac{f_1+f_{2}}{2}\right)^{2}}+\frac{\Im(s_2^+)^{2}}{(d -f_1)(d - f_2)}<1$, while the outer region $\frac{\left(\Re(s_2^+)-f_{2} / 2\right)^{2}}{\left(d-\frac{f_1+f_{2}}{2}\right)^{2}}+\frac{\Im(s_2^+)^{2}}{(d -f_1)(d - f_2)}>1$ is may not be analytic. 
\\
The above curve \eqref{eq:masterdomain} has three different choice of $s_1$ listed in \eqref{eq:martin} . We will name the curves as follows
\be\label{eq:rangeeq}
\begin{split}
&C_1(s_1): \text{ is the curve \eqref{eq:masterdomain} for } \frac{2\m}{3}<s_1<\frac{11\mu}{3}\\
&C_2(s_1): \text{ is the curve \eqref{eq:masterdomain} for } \frac{11\m}{3}<s_1<\frac{23\mu}{3}\\
&C_3(s_1): \text{ is the curve \eqref{eq:masterdomain} for } \frac{23\m}{3}<s_1<\infty
\end{split}
\ee 

\subsection{Overall domain drawn out of the curve $C_1(s_1)$}
\begin{figure}[h!]
  \centering
  \begin{subfigure}[b]{0.45\linewidth}
    \includegraphics[width=\linewidth]{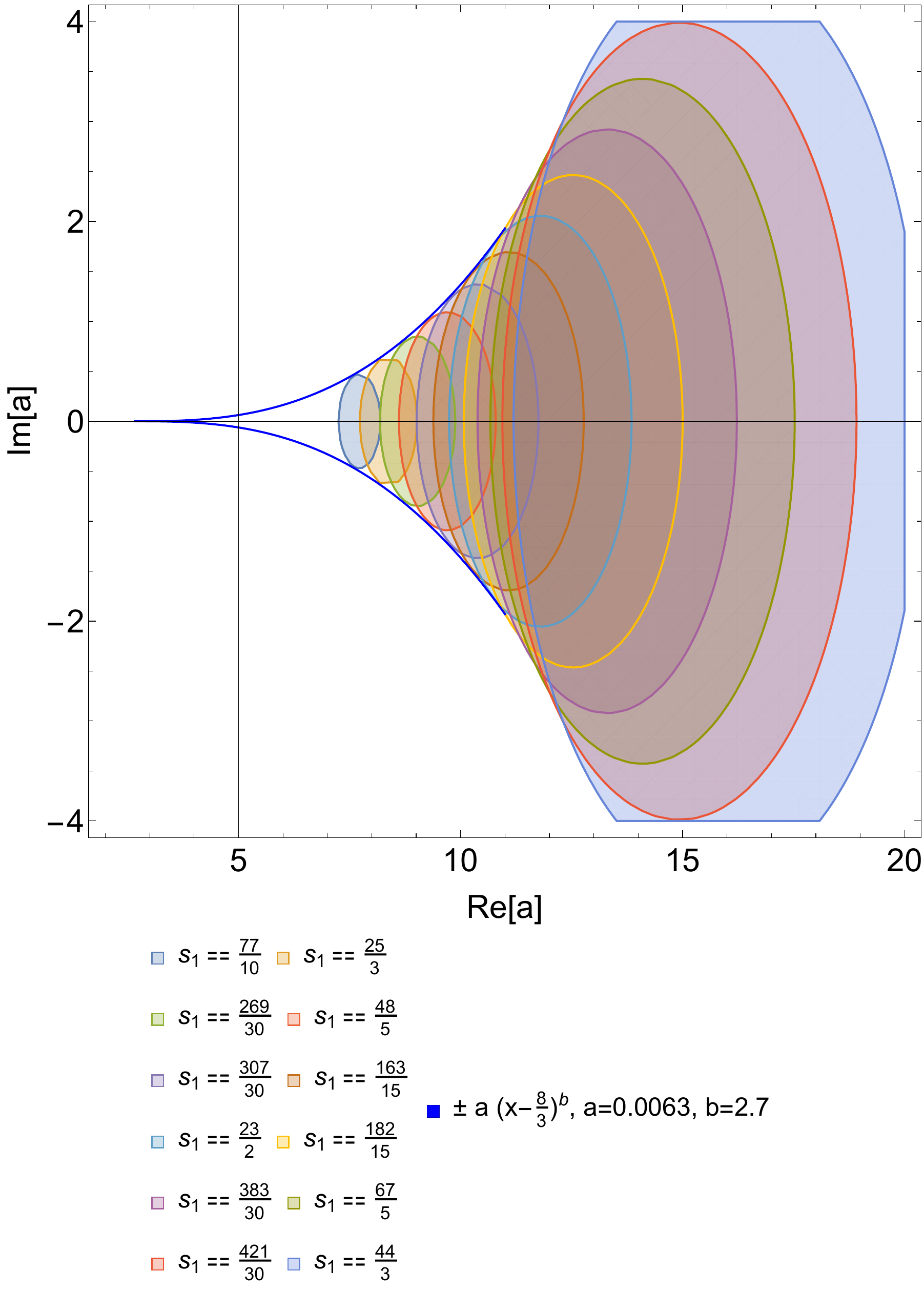}
\caption{The analyticity domain in complex $a$-plane for various $s_1$ in the range $\frac{2 \mu }{3}<s_1<\frac{11 \mu }{3}$.}
\label{fig:region1}
  \end{subfigure}
  \begin{subfigure}[b]{0.45\linewidth}
    \includegraphics[width=\linewidth]{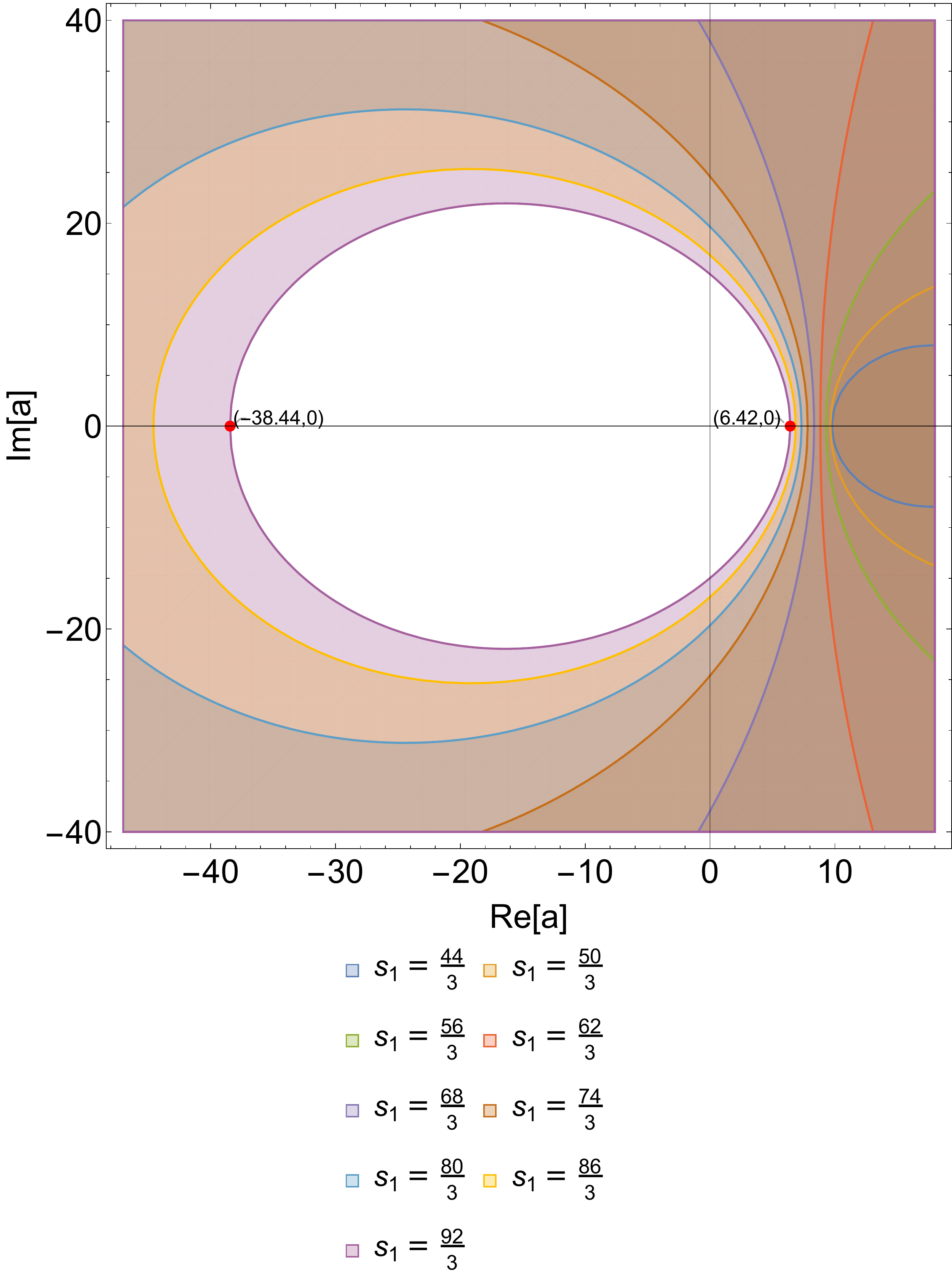}
\caption{The analyticity domain in complex $a$-plane for various $s_1$ in the range $\frac{11 \mu }{3}<s_1<\frac{23 \mu }{3}$.}
\label{fig:region2}
  \end{subfigure}
  \begin{subfigure}[b]{0.45\linewidth}
    \includegraphics[width=\linewidth]{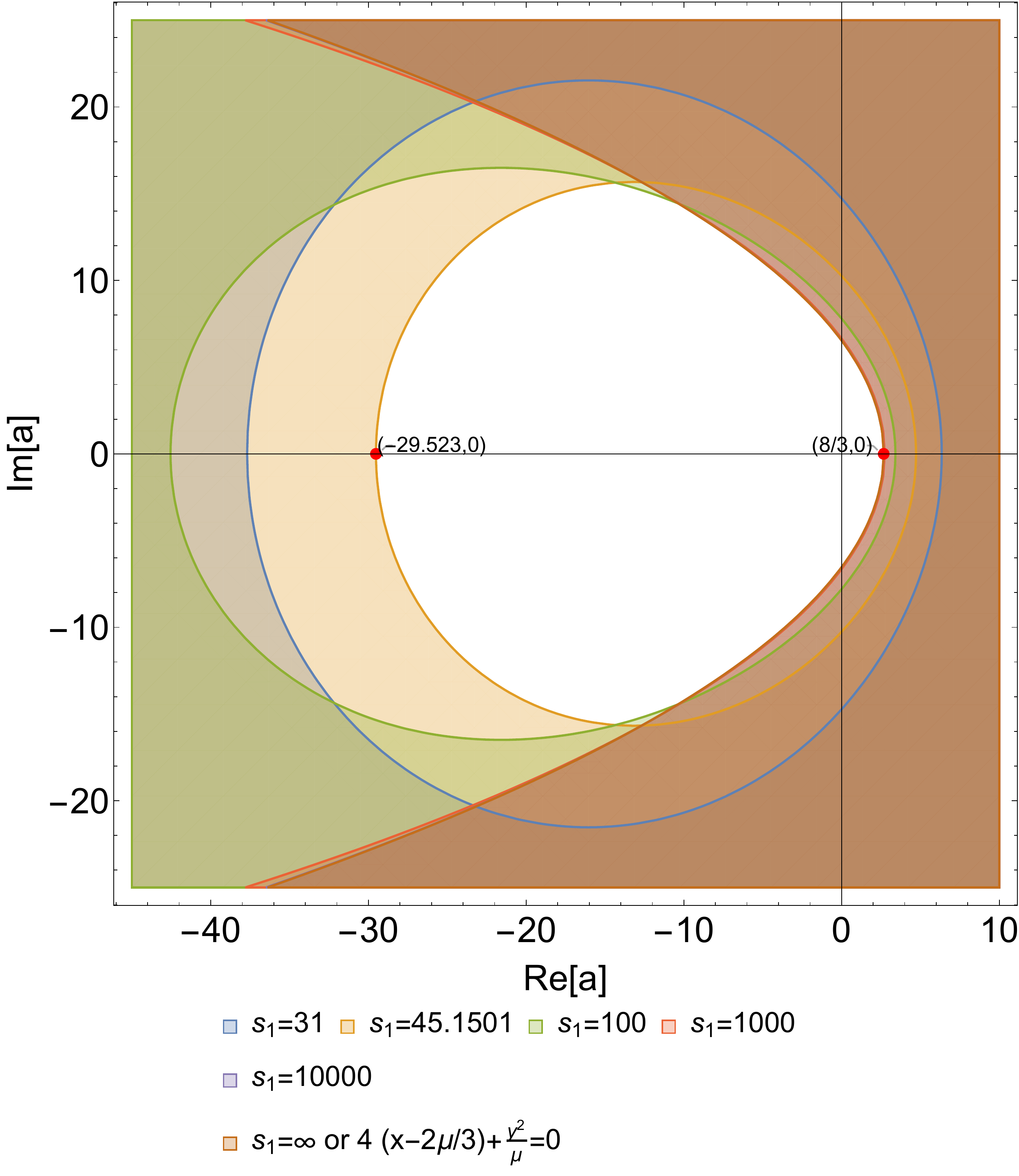}
\caption{The analyticity domain in complex $a$-plane for various $s_1$ in the range $\frac{23 \mu }{3}<s_1<\infty$.}
\label{fig:region3}
  \end{subfigure}
  \begin{subfigure}[b]{0.45\linewidth}
    \includegraphics[width=\linewidth]{range_a_all}
\caption{The analyticity domain in complex $a$-plane for various $s_1$ in the range $\frac{2 \mu }{3}<s_1<\infty$.}
\label{fig:allrange}
  \end{subfigure}
  \caption{The analyticity domain in complex $a$-plane. Shaded region is excluded.}
  \label{fig:coffee3}
\end{figure}

For the case $\frac{2 \mu }{3}<s_1<\frac{11 \mu }{3}$, each curve $C_1(s_1)$ looks similar to an ellipse. The area outside of these curves are analytic regions. In the figure \eqref{fig:region1} the curves are shown for various values of $s_1$.

The shaded regions may not be analytic, while the outside white region is analytic.  As we decrease $s_1$ from $\frac{11\mu}{ 3}$ to $\frac{2\mu}{ 3}$, we see that effective area of $C_1(s_1)$ decrease. As we evolve the $C_1(s_1)$ from $\frac{11\mu}{ 3}$ to $\frac{2\mu}{ 3}$, the maxima traces two curves. These two curves end at $s_1=2\m/3$ in the real axis $\Re(a)=2\m/3$. For $\mu=4$, we find the fit $\pm a \left(x-\frac{8}{3}\right)^b\,,$ with   $a=0.0063,b=2.65$ for $x>8/3$ and $s_1$ is close to $8/3$. In the figure, these fit curves are presented as two blue line. Then overall region of analyticity will be outside of these two curve formed by movement of the $C_1(s_1)$ from  $\frac{11\mu}{ 3}$ to $\frac{2\mu}{ 3}$. This is shown in the figure \eqref{fig:region1}. The white region is analyticity domain of the scattering amplitude in the complex $a$ plane for $\frac{2 \mu }{3}<s_1<\frac{11 \mu }{3}$.

\subsection{Overall domain drawn out of the curve $C_2(s_1)$}
For the case $\frac{11 \mu }{3}<s_1<\frac{23 \mu }{3}$, each curve $C_2(s_1)$ are shown in the figure \eqref{fig:region2} for various values of $s_1$. The area inside of these curves are the analytic regions for the scatting amplitude in the complex $a$ plane. The shaded regions are may not be analytic, while the white region is analytic.
The overall region of analyticity is formed by the curve $C_2(s_1=\frac{23 \mu }{3})$. For the case of $\m=4$, the end point in real axis are indicated by red dots in the plot.

\subsection{Overall domain drawn out of the curve $C_3(s_1)$}

For the case $\frac{23 \mu }{3}<s_1<\infty$, each curve $C_3(s_1)$ are shown in the figure \eqref{fig:region3} for various values of $s_1$. The area outside of these curves are analytic regions for scattering amplitude in  the complex $a$-plane. The shaded regions are may not be analytic, while the white region is analytic.
The overall region of analyticity is formed by the curve $C_3(s_1=\infty)$ and the optimal curve which touch the real axis such that $-\Re[a]$ is minimum (We will provide explicit calculations in a moment). For the case of $\m=4$, the end point in real axis are indicated by red dots in the plot.

\subsubsection*{The equation for the curve $C_3(s_1=\infty)$}
The equation for the curve $C_3(s_1=\infty)$ can be explicitly worked out. We can expand \eqref{eq:masterdomain} for large $s_1$, which is
\be
1+\frac{4(\Re(a)-2\mu/3)+\Im(a)^2/\mu}{s_1}+O\left(\frac{1}{s_1^2}\right)=1
\ee
For large $s_1$, we have the equation
\be
4\left(\Re(a)-\frac{2\mu}{3}\right)+\frac{\Im(a)^2}{\mu}=0
\ee
Note that analyticity domain is $4(\Re(a)-2\mu/3)+\frac{\Im(a)^2}{\mu}<0$, while $4(\Re(a)-2\mu/3)+\frac{\Im(a)^2}{\mu}>0$ is excluded.

\subsubsection*{The optimal curve $C_3(s_1)$ which touches the real axis such that $-\Re(a)$ is minimum.}
Now we want to find the optimal curve which touch the real axis such that $-\Re[a]$ is minimum. Note that for real $a$, we have $\Im[s_2]=\frac{1}{2} s_1\sqrt{\frac{-3 a-s_1}{s_1-a}}$. Note that equation \eqref{eq:masterdomain} implies that analyticity domain is determined $\Im[s_2]<B$. The optimal value of $s_1$ determined by $\Im[s_2]=B$, which minimum the value of $-a$ on the real axis, explicitly 
\be
\frac{1}{2} s_1 \sqrt{\frac{-3 a-s_1}{s_1-a}}=\sqrt{\left(\mu +\frac{48 \mu }{3 s_1-11 \mu }\right) \left(\frac{\mu }{3}+\frac{48 \mu }{3 s_1-11 \mu }+s_1\right)}
\ee
We will work with $\m=4$. The equation gives 
\be
a=\frac{-27 s_1^5+360 s_1^4-624 s_1^3-11520 s_1^2-25600 s_1}{81 s_1^4-2808 s_1^3+22608 s_1^2-11520 s_1-25600}
\ee
Now we can minimize $-a$ from the above equation with respect to $s_1$. The value of $s_1=45.1501$ minimizes $-a$ and the value determined is  $a=-29.523$. This value is indicated in figures \eqref{fig:region3} and \eqref{fig:allrange}.

\subsection{Overall domain drawn out of the curves $C_1(s_1),~C_2(s_1),~C_3(s_1)$}
We can combine all three regions, the overall domain of analyticity is shown in the figure \eqref{fig:allrange}.
The overall domain is controlled by the three curves. These are 
\begin{enumerate}
\item $4\left(\Re(a)-\frac{2\mu}{3}\right)+\frac{\Im(a)^2}{\mu}=0$. This is the curve $C_3(s_1=\infty)$.
\item Two curves traced by evolution of the maxima of $C_1(s_1)$ by varying   $s_1$ from $\frac{11\mu}{ 3}$ to $\frac{2\mu}{ 3}$. For $\mu=4$, we find the fit $\pm a \left(x-\frac{8}{3}\right)^b\,,$ with   $a=0.0063,b=2.65$ for $x>8/3$ and $s_1$ is close to $8/3$.
\item The optimal $C_3(s_1)$ curve which touches the real axis such that $-\Re(a)$ is minimum. For $\m=4$, we find $s_1=45.1501$ with $-\Re(a)_{min}=29.523$
\end{enumerate}
In our notation\footnote{Note that $a_{their}-\frac{\mu}{3}=a_{our}$} the analyticity domain for real $a$, we have $-29.523<a<8/3$ for $\mu=4$, which is consistent with \cite{AK}.

\section{Analyticity domain for Dyson block and Feynman block}
\begin{figure}[hbt!]
  \centering
  \begin{subfigure}[b]{0.45\linewidth}
    \includegraphics[width=\linewidth]{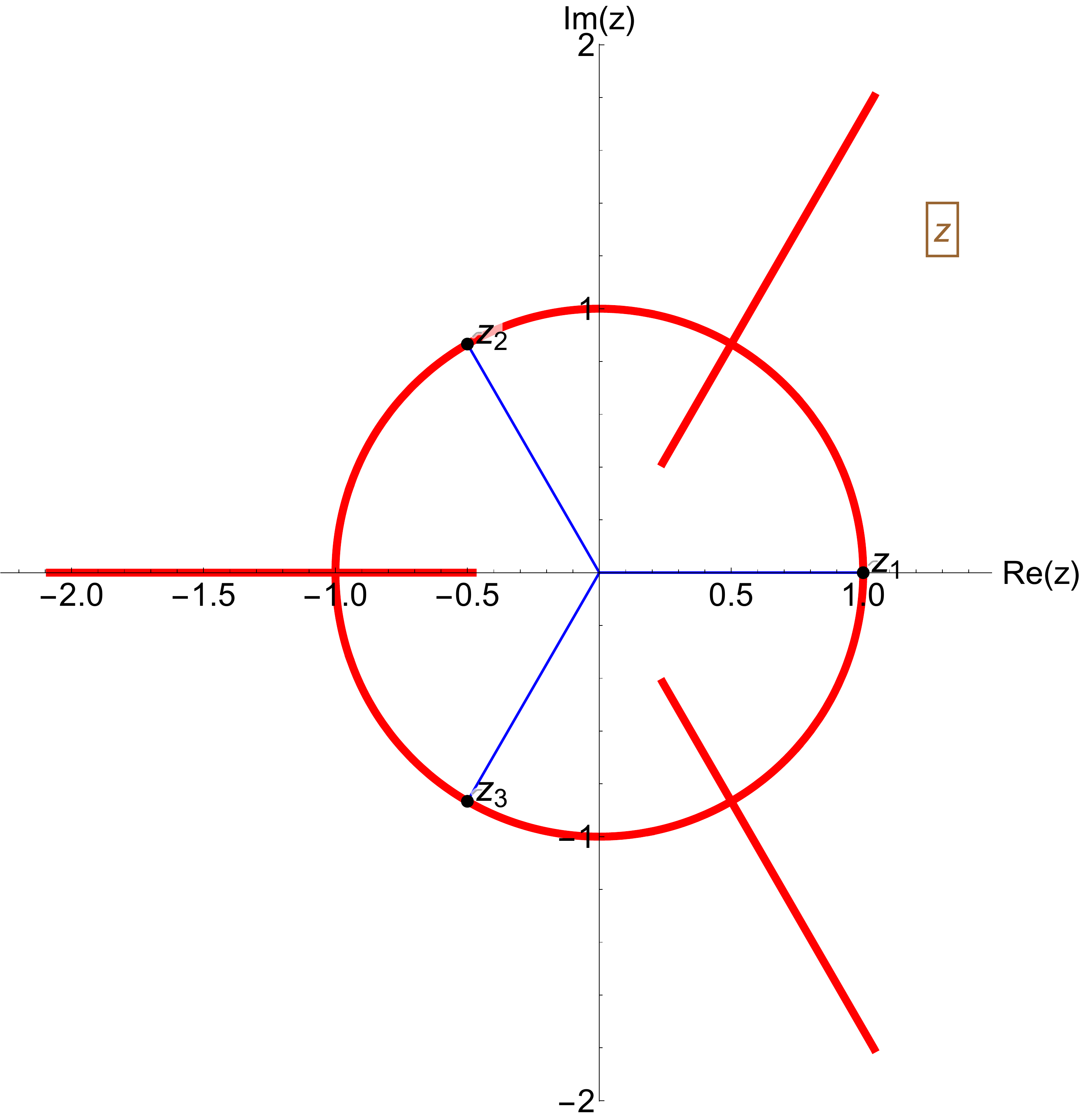}
\caption{Image of the physical cuts for case-III \textit{i.e.} real $a<-2\mu/9$. We choose $a=-8/9-1/2$.}
\label{fig:figz1}
  \end{subfigure}
  \begin{subfigure}[b]{0.45\linewidth}
    \includegraphics[width=\linewidth]{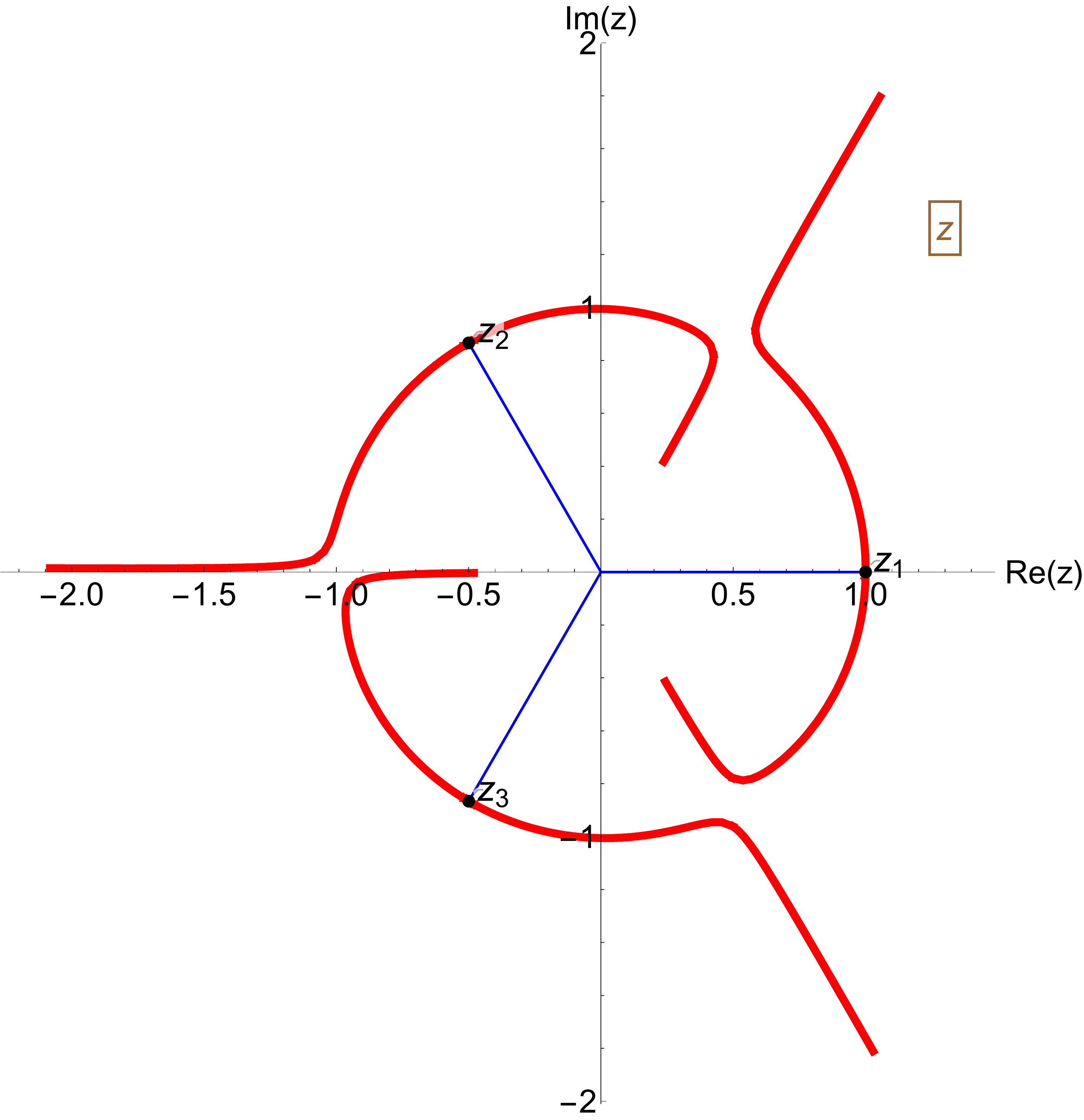}
\caption{Image of the physical cuts for complex $a$. We choose $a=-8/9-1/2+i/100$.}
\label{fig:figz2}
  \end{subfigure}
  \caption{Image of the physical cuts in the complex $z$-plane. We choose $\m=4$ for the above plots. }
  \label{fig:figzcuts}
\end{figure}

{ Now we would like to investigate the analyticity domains for Dyson and Feynman block expansions for the \emph{scattering amplitude}. Recall that the Dyson block expansion is obtained by inserting the usual partial wave expansion into CSDR. The Lehmann-Martin ellipses are domains of analyticity for the absorptive part. When translated in terms of the variable $a$, while keeping $s_1$ unchanged, \cite{AK} established that for the real segment of $a$ contained in the intersections of the Lehamnn-Martin ellipses, $a\in (-29.523, 8/3)$, analytic structure of the  CSDR kernel established that amplitude itself was analytic as well, thereby establishing the convergence of the Dyson block expansion in the same domain. The non-trivial part behind this conclusion is that it takes into account the \emph{full crossing symmetry in all the three channels simultaneously} which was \emph{not} used in the original derivation \cite{Martinlec} of the Lehmann-Martin elliptical domains.  
	
	Due to some technical subtlety, \cite{AK} did not extend the analysis to complex domains in $a$.{ We explore the problem of analyzing and deriving the analyticity in complex $a$-domain, and \emph{numerically} we illustrate our domain of validity for some particular amplitudes}. Along with this, we take up a similar exploration for Feynman block expansion. Further, we make a comparison of the convergence properties of the Feynman block expansion with that of the Dyson block expansion. While both of these expansion blocks are manifestly crossing-symmetric, the Feynman blocks are manifestly local, and therefore the point of emphasis of our comparison is the \emph{possible consequence of the locality} on the convergence properties.   
\subsection{Expectation for the analyticity domains}
We see that the CSDR contains an integral over the entire physical domain of $s_1$, $2\m/3\leq s_1<\infty$. Thus we need to have a common domain of analyticity for the absorptive part for the entire physical range of $s_1$ to begin  with. This domain will be the common intersection of the Lehmann-Martin ellipses \eqref{eq:martin}. We have analyzed this intersection in the previous section.  To recapitulate, this overall domain will be controlled by the two curves. These are 
\begin{enumerate}
\item $4\left(\Re(a)-\frac{2\mu}{3}\right)+\frac{\Im(a)^2}{\mu}=0$. This is the curve $C_3(s_1=\infty)$.
\item The optimal $C_3(s_1)$ curve which touch the real axis such that $-\Re(a)$ is minimum. For $\m=4$ we find $s_1=45.1501$ with $-\Re(a)_{min}=29.523$
\end{enumerate}
While this domain guarantees analyticity of the absoprtive part over the entire range of $s_1$ integration, the same for the rest of the CSDR integral is not obvious at all! While this has been established for the real range of $a$ belonging to this common domain in \cite{AK}, {We find that similar argument holds for the above domain in complex $a$ plane, and numerical investigations support our findings}. First we look for whether the Dyson block expansion, i.e. the entire integral converges, in this domain for a few specific \emph{non-trivial} amplitudes. Then  we carry out the same exercise for the Feynman block expansion. We would like to emphasize that in both of these investigations, we are looking for \emph{actual} convergence domains for the corresponding block expansions which \emph{may or may not} be the common domain obtained from the Lehmann-Martin ellipses as mentioned above. 

\subsection{Why analyticity in complex $a$ domains is important for CSDR?}

Let us discuss one important consequence of complex $a$ values for the CSDR. In the derivation of the CSDR in \cite{AK,ASAZ}, only real values of $a$ were considered. For real values of $a$ the physical cuts map to the unit circles or arc of unit circles in the complex $z$-plane. The authors divided the real range of $a$ into three cases. For case-I the range is $-2\mu/9<a<0$, for case-II, the range is $0<a<2\mu/3$ and for case-III, the range is $a<-2\mu/9$. The cases-I,II show gap in the unit circles or arcs of the unit circles in the complex $z$-plane that represents the physical cuts. Therefore one can easily analytically continue from inside the unit-circle to outside the circle. This analytic continuation is crucial for writing down the dispersion relation. For detailed discussion about the cases I,II and derivation of the dispersion relation, we refer the reader to the appendix of \cite{ASAZ}. But naive analytic continuation from inside the unit-circle to outside the circle is not possible for case-III. This is because the arcs join up for real values of $a$. For example see figure \eqref{fig:figz1}. This kind of analytic continuation can be done after we consider complex values of $a$. See for example figure \eqref{fig:figz2}. Slight complex shift in $a$ opens up the unit circle. Therefore the reason  the dispersion works for case-III is clear now. \emph{It picks up a complex path in the complex $a$-plane to make analytic continuation from inside the unit-circle to outside the circle.}
}

\subsection{Numerical illustrations}
Now that the basic ground for our numerical analysis has been laid out, we demonstrate our findings of contact terms, analyticity domain of Feynman and Dyson block for tree level type-II superstring  amplitude and Pion S-matrices. 
\subsubsection*{Type II superstring tree amplitude}
\begin{figure}[b!]
  \centering
  \begin{subfigure}[b]{0.43\linewidth}
    \includegraphics[width=\linewidth]{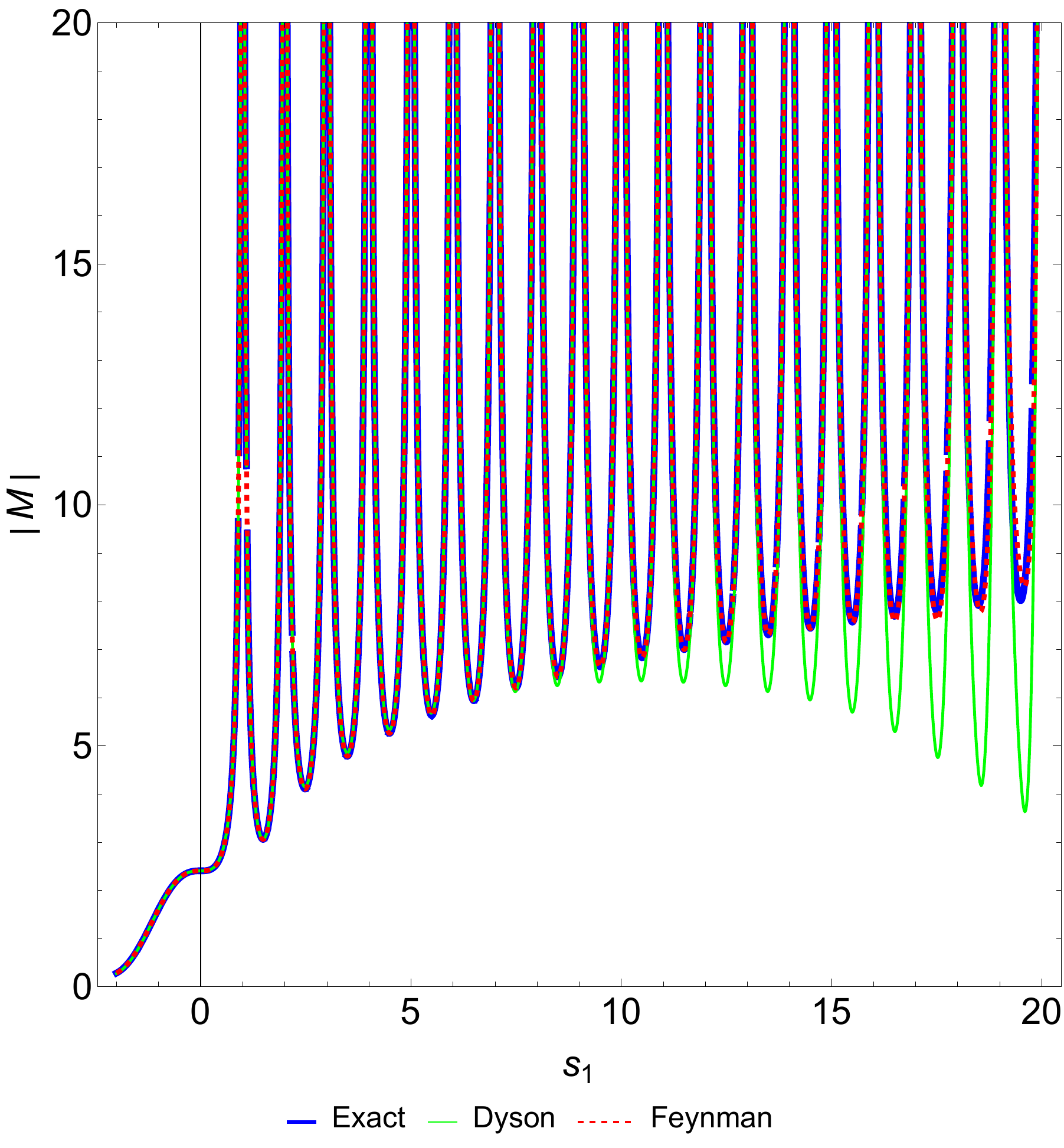}
\caption{For fixed $a=\frac{11}{10}+\frac{9}{10} i$, we compare the exact amplitude and the Dyson and Feynman block expansion as a function of $s_1$ with $k_{\max }=30, L_{\max }=30$. Blue curve is exact, green is Dyson and red dashed is Feynman block expansion.}
\label{fig:string_fixa}
  \end{subfigure}
  \begin{subfigure}[b]{0.45\linewidth}
    \includegraphics[width=\linewidth]{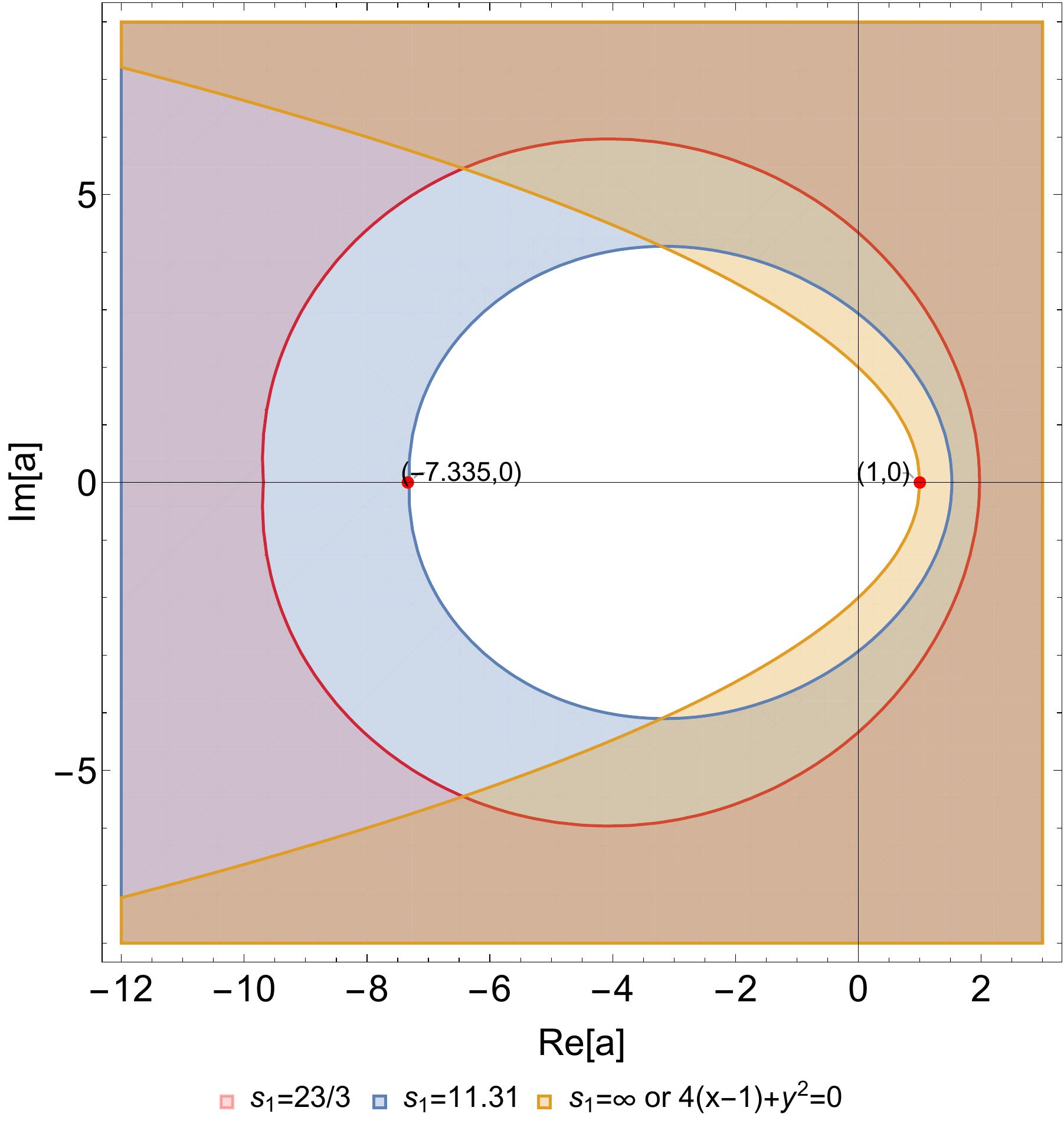}
\caption{The analyticity domain in complex $a$-plane for various $s_1$ in the range $\d_0 <s_1<\infty$. We consider $s_1=23/3$ for various examples is shown in red contour. Other two curves dictate the boundary of the overall domain.}
\label{fig:string_reg}
  \end{subfigure}
  \begin{subfigure}[b]{0.45\linewidth}
    \includegraphics[width=\linewidth]{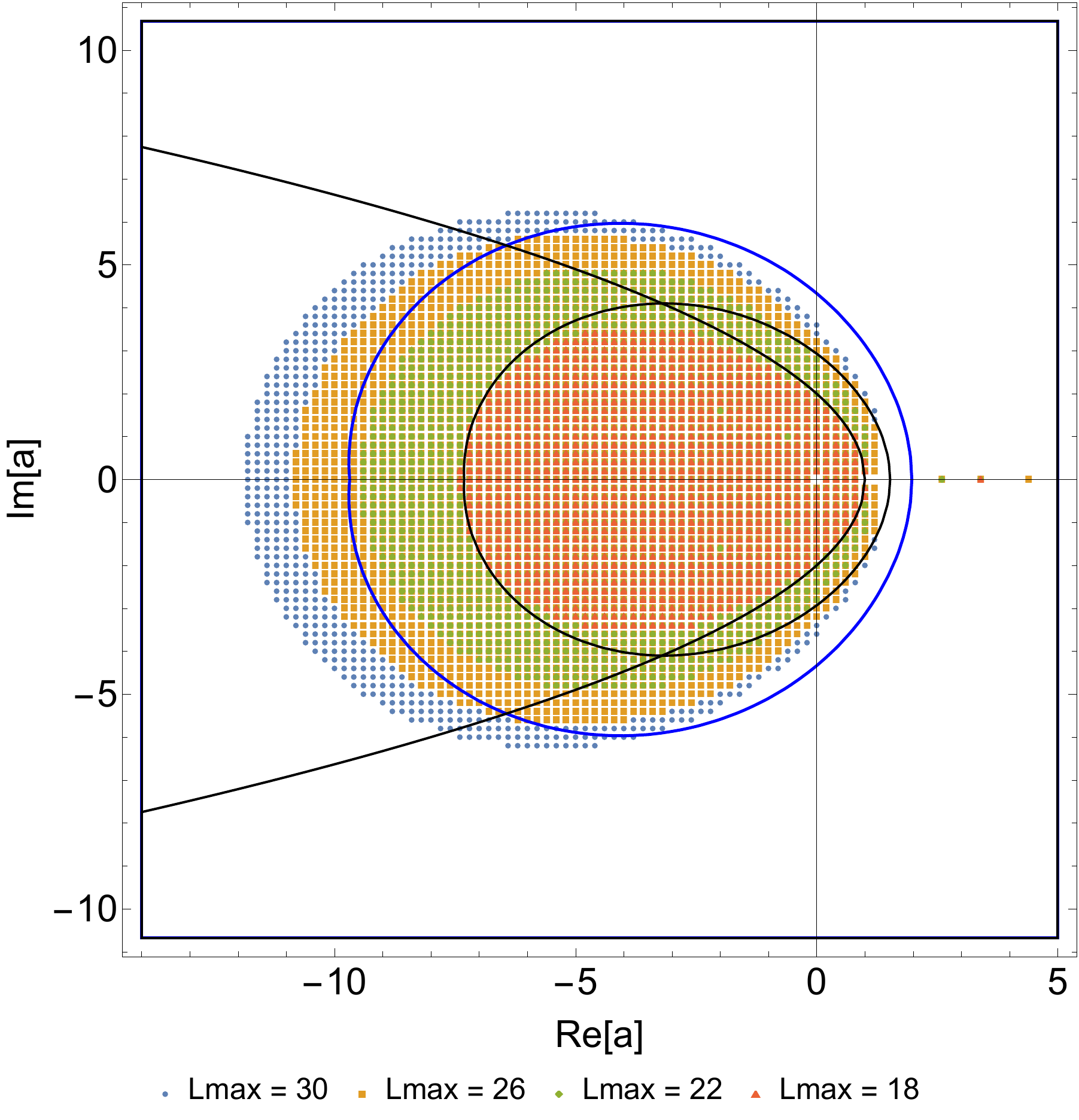}
\caption{Dyson block expansion of amplitude with $k_{\max }=40$. We consider $s_1=23/3$, shown in blue contour and black lines are boundary of the overall domain.}
\label{fig:string_dyson}
  \end{subfigure}
  \begin{subfigure}[b]{0.45\linewidth}
    \includegraphics[width=\linewidth]{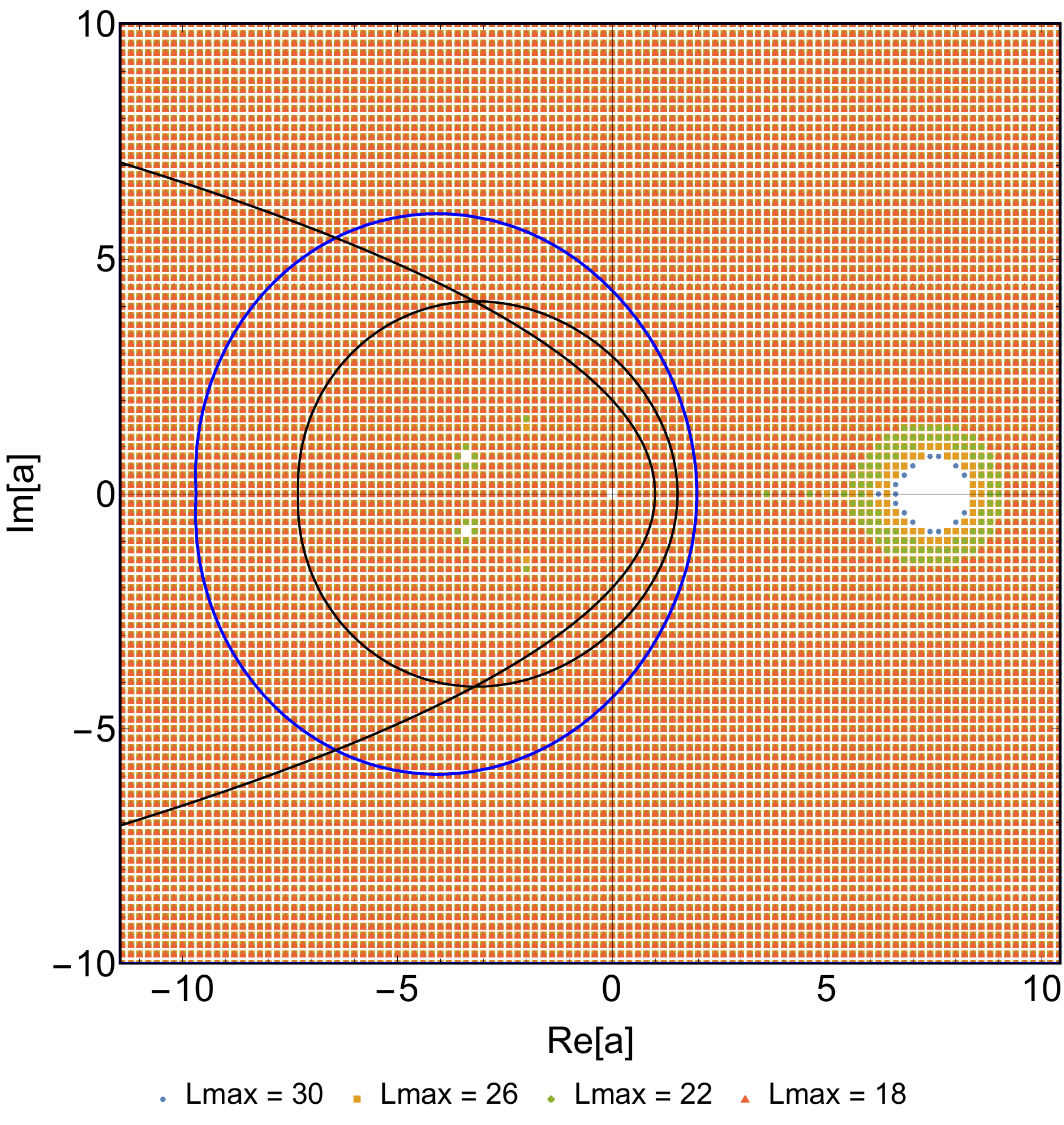}
\caption{Feynman block expansion of amplitude with $k_{\max }=40$. We consider $s_1=23/3$, shown in blue contour and black lines are boundary of the overall domain.}
\label{fig:string_feynman}
  \end{subfigure}
  \caption{Numerical illustrations using string amplitude.}
  \label{fig:string_demo}
\end{figure}
We consider the four dilaton type II superstring tree amplitude (see for example \cite{green}) for our illustrations. We will subtract out the massless pole $\frac{1}{s_1 s_2 s_3}$.
\be
\mathcal{M}\left(s_{1}, s_{2}\right)=-\frac{\Gamma\left(-s_{1}\right) \Gamma\left(-s_{2}\right) \Gamma\left(s_{1}+s_{2}\right)}{\Gamma\left(s_{1}+1\right) \Gamma\left(-s_{1}-s_{2}+1\right) \Gamma\left(s_{2}+1\right)}+\frac{1}{s_1 s_2 (s_1+s_2)}
\ee
Above amplitude is massless external but has a gap $\d_0=1$. For such cases we presume that equation for Lehmann-Martin ellipse is give by $E\left(0, \d_0-s_1 \mid \d_0 +\frac{4 \d_0 ^2}{s_1-4 \d_0 }\right)$, where $E\left(f_{1}, f_{2} \mid d\right)$ stands for an ellipse with foci at $s_{2}^{+}=f_{1}, s_{2}^{+}=f_{2}$. See figure \eqref{fig:string_reg}. The absorptive part is given by $\mathcal{A}\left(s_{1} ; s_{2}\right)=$ $\sum_{k=1}^{\infty} \frac{(-1)^{k+1} \Gamma\left(-s_{2}\right) \Gamma\left(k+s_{2}\right)}{(k !)^{2} \Gamma\left(s_{2}+1\right) \Gamma\left(-k-s_{2}+1\right)} \pi \delta\left(s_{1}-k\right)$. We shall work in four space-time dimensions, i.e., $\alpha=1 / 2$. The partial wave coefficients are given by
$$
\Phi(\sigma ; 1 / 2) a_{\ell}(\sigma)=\sum_{k=1}^{\infty} \pi \delta(\sigma-k) \frac{1}{2} \int_{-1}^{1} d x \frac{(-1)^{k+1} \Gamma\left(-\frac{1}{2} k(x-1)\right) \Gamma\left(\frac{1}{2} k(x+1)\right)}{(k !)^{2} \Gamma\left(\frac{1}{2} k(x-1)+1\right) \Gamma\left(1-\frac{1}{2} k(x+1)\right)} C_{\ell}^{(1 / 2)}(x)
$$
We do $\sigma$ integral by picking up $\sigma=k$ from the $\delta$-function. We can put this in the Dyson and Feynman block expansion. Numerical illustrations is shown in the figure \eqref{fig:string_demo}.

We can fix $a$ then compare the exact amplitude and the Dyson and Feynman block expansion as a function of $s_1$. For numerical purpose, we truncate the $k, \ell$ sum upto $k_{\max }=30, L_{\max }=30$. See figure \eqref{fig:string_fixa} for fix $a=\frac{11}{10}+\frac{9}{10} i$. One can easily see that Feynman block is of better convergent.

We define the error function $err=\left| \frac{M_{exact}-M_{D/F}}{M_{exact}}\right|$, where $M_{D/F}$ stands for Dyson or Feynman block expansion of the amplitude. We fix the $err$ to be $10\%$ or simply $err=0.1$ and observe the convergence in the complex $a$-plane. For $10\%$ accuracy as we increase the spin we see that region of convergence grows with the spins. See figure \eqref{fig:string_dyson} for Dyson block expansion and figure \eqref{fig:string_feynman} for Feynman block expansion. Again we can see that the Dyson block expansion is convergent and almost contained within the Lehmann-Martin ellipse. If one increases spins, it goes beyond the ellipse. In contrary the Feynman block covers almost the whole complex plane.

\subsubsection*{Pion S-matrices}
\begin{figure}[hbt!]
  \centering
  \begin{subfigure}[b]{0.4\linewidth}
    \includegraphics[width=\linewidth]{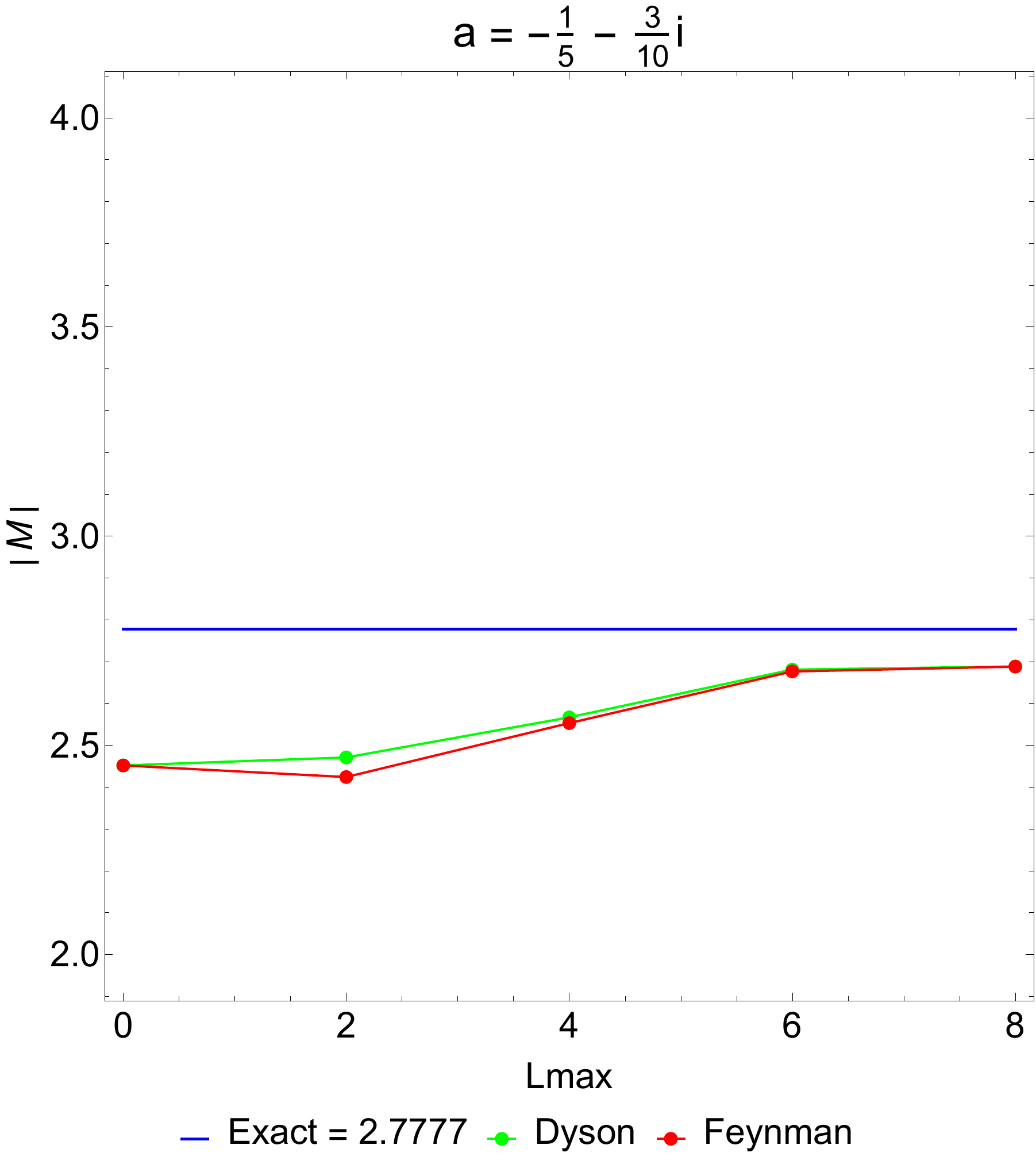}
\caption{}
\label{fig:pion1}
  \end{subfigure}
  \begin{subfigure}[b]{0.4\linewidth}
    \includegraphics[width=\linewidth]{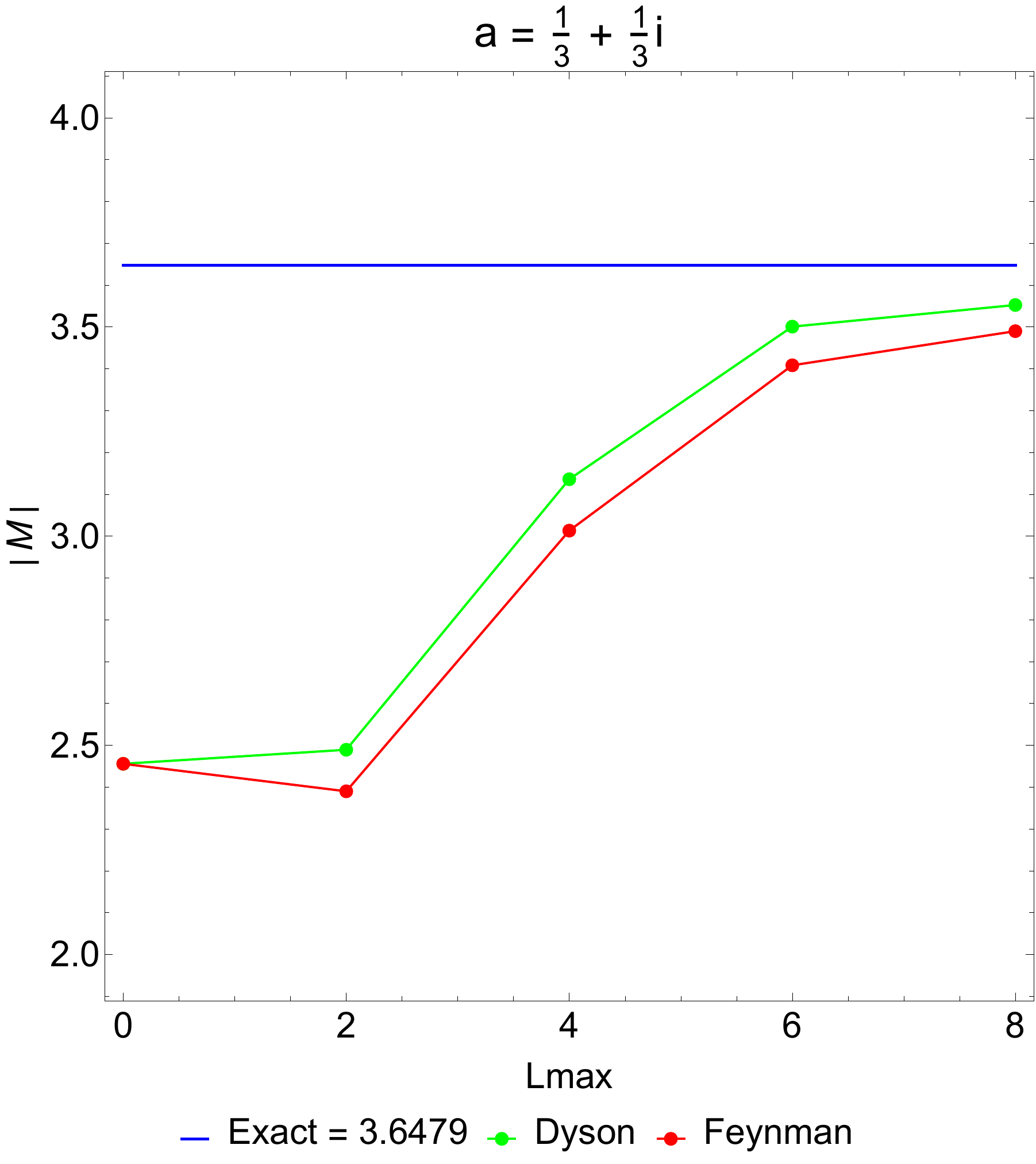}
\caption{ }
\label{fig:pion2}
  \end{subfigure}
  \caption{Numerical illustrations using pion amplitude in \cite{ABAS, ABPHAS}. Dyson block and Feynman block expansion of amplitude. We consider $s_1=111/10$.}
  \label{fig:pion_demo}
\end{figure}

We present numerical illustrations of the Dyson and Feynman block expansion of amplitude using pion amplitude (for $s_0=0.36, s_2=2.08$ in the upper river boundary) in \cite{ABAS, ABPHAS} in figure \eqref{fig:pion_demo}. We consider $s_1=111/10$. Values of the $a$ indicated in the figures. We see that both Dyson and Feynman block expansions are convergent for complex $a$. The basis expansion presented in \cite{ABAS,ABPHAS} is quite difficult to handle in numerics we are doing, specially scanning the complex $a$-plane as shown for string amplitude case, since this type of scanning requires larger $L_{max}$.

 { \subsection{Observations : Reading between the lines}
{From our numerical investigations, we find that  the convergence domain in the complex $a$-plane for Dyson block expansion exactly coincides with the analogues of the Lehmann-Martin ellipses}. Further, the analysis of four dilaton amplitude in type-II superstring theory  gives us that the Feynman block expansion has a \emph{bigger} convergence domain than that of the Dyson block expansion. Since Feynman block emerges after imposing the locality constraints,  \emph{a priori} it's not clear why this should be the case.  A complete proof will require a systematic understanding of various locality constraints as well as the newly derived contact terms, \eqref{eq:Feynmancm}, towards the convergence properties. In a way the numerical investigations that we took up prepare the ground for the more consuming task of providing robust analytic support to these various results!  }


\section{Summary and future directions}
In this paper, we discuss locality and analyticity domain of crossing symmetric dispersion relation. The main results of the paper are the following:
\begin{itemize}
\item Imposing locality constraints on the crossing symmetric dispersion relation \eqref{AKDR} give us a fully crossing symmetric and local expansion of scattering amplitudes. Crossing symmetric dispersion relation has non-local terms, which should be vanished. These non-local terms can be removed systematically. After we remove these terms, we get Feynman block expansion. \textit{We have written down the general formula of such Feynman block expansion \eqref{eq:Feynmanstu} deriving the completely general form of the contact terms \eqref{eq:Feynmancm}.}


\item \textit{We investigate the analyticity domain of the crossing symmetric dispersion relation \eqref{AKDR} in complex $a$ plane} analogous to Lehmann-Martin ellipses. See figures \eqref{fig:region1}, \eqref{fig:region2}, \eqref{fig:region3}, \eqref{fig:allrange}. We further investigate the \emph{analyticity domain} for the Feynman block expansion, \eqref{fig:allrange}. 

\item We have supported our findings by comparing them against know theories.  \textit{We have shown that pion S-matrices found in \cite{ABPHAS,ABAS} show good agreement with the Feynman block expansion (see figure \eqref{fig:pion_demo}).}

\end{itemize}

Below we tabulate some of the immediate future goals. 
\begin{itemize}

\item \textbf{More on Feynman blocks:} We have only uncovered tip of an iceberg on Feynman block expansions. On the technical front, we need to systematically analyze the convergence properties of the Feynman block expansion analytically, and possibly give an analytic derivation of  the numerical observations obtained in this work.  On a more conceptual ground, the evidence of the Feynman block expansion having larger convergence domains ignites the possibility of  its connection with  Mandelstam analyticity! In particular, it will be worthwhile to understand the precise role of the locality upon the analyticity properties of the scattering amplitude. Further, since we have now explicit analytic expressions for the Feynman blocks ( contact terms), it is possible to explore multiple avenues. For example, it will be interesting to further explore connection with the geometric function theoretic insights recently uncovered  in \cite{SGASPR}. 

\item \textbf{CFT applications:} The analysis that we partook in this work can be employed to Mellin amplitudes in conformal field theory quite strainghtforwardly. In particular, we present the derivation of the general formula for the contact terms for CFT Mellin amplitudes in near future.Using the general formula of the contact term, in principle one can find the $\e^3$ anomalous dimension for $\D_{\phi^2}$ operator using Polyakov-Mellin Bootstrap or similar higher order corrections untreated in \cite{Pol, ks, usprl, ON, PFKGASAZ} due to lack of contact terms. Further  it would be interesting to find out the rigorous analyticity domain for Witten block expansion of the CFT Mellin amplitudes. Another exciting application will be to explore our discussion in the context of the CFT correlators of charged fields discussed in \cite{Ghosh:2021ruh} and the functionals therein also to \cite{Apratim, Alday}. It will be fascinating to co-relate our Feynman block expansion or Witten block expansion with swampland conditions considered recently in  as well as the correlator bounds in \cite{Paulos:2020zxx}. It will be very interesting to relate positivity of Feynman block or Witten block analysis to the positive geometries discussed in \cite{positivegeo}.

\item \textbf{Numerical S-matrix bootstrap:} Another crucial application will be to set up the S-matrix bootstrap using the Feynman block as done for S-matrix bootstrap for pion amplitudes discussed \cite{ABPHAS,ABAS, andrea} and for the dual S-matrix bootstrap \cite{dual}. Since Feynman block is local and fully crossing symmetric it will be more powerful. 

\item \textbf{Spinning amplitudes:} On a more technical front, it will be intriguing to explore our discussion in the context of spinning particles \cite{spinning} as well as \cite{Bern:2022yes, Multi, Rastelli,Karateev:2022jdb}. It will be worth to explore recent ideas \cite{Moments, Causality, Reverse, Yutinnew1} in the context of Feynamn block expansion.

\end{itemize}

\section*{Acknowledgments}

We thank Aninda Sinha for suggesting the problem, his constant encouragement and numerous helpful discussions. We thank Faizan Bhat, Sudip Ghosh, Prashanth Raman for their comments on the draft. We also thank Shaswat Tiwari for helping us with the S-matrix bootstrap data.

\appendix

\section{Lehmann-Martin ellipses}\label{sec:Lehmann-Martin ellipses}

The analyticity domain in complex $s_2$ for various range of $s_1$ is given by \eqref{eq:martin}, which are Lehmann-Martin ellipses. The equation in the complex $s_2$ plane is given by
\be
\label{eq:martindomain}
\frac{\left(\Re(s_2)-f_{2} / 2\right)^{2}}{\left(d-\frac{f_1+f_{2}}{2}\right)^{2}}+\frac{\Im(s_2)^{2}}{(d -f_1)(d - f_2)}=1
\ee

We have the analyticity domain in complex $s_2$ plane which governed by the above equation of ellipse. The domain analyticity domain is $\frac{\left(\Re(s_2)-f_{2} / 2\right)^{2}}{\left(d-\frac{f_1+f_{2}}{2}\right)^{2}}+\frac{\Im(s_2)^{2}}{(d -f_1)(d - f_2)}<1$, while the region outside $\frac{\left(\Re(s_2)-f_{2} / 2\right)^{2}}{\left(d-\frac{f_1+f_{2}}{2}\right)^{2}}+\frac{\Im(s_2)^{2}}{(d -f_1)(d - f_2)}>1$ is may not be analytic. 
\\
The above equation \eqref{eq:martindomain} has three different choice of $s_1$ listed in \eqref{eq:martin}. We will name the curves as follows
\be
\begin{split}
&LM_1(s_1): \text{ is the ellipse \eqref{eq:martindomain} for } \frac{2\m}{3}<s_1<\frac{11\mu}{3}\\
&LM_2(s_1): \text{ is the ellipse \eqref{eq:martindomain} for } \frac{11\m}{3}<s_1<\frac{23\mu}{3}\\
&LM_3(s_1): \text{ is the ellipse \eqref{eq:martindomain} for } \frac{23\m}{3}<s_1<\infty
\end{split}
\ee 

\begin{figure}[hbt!]
  \centering
  \begin{subfigure}[b]{0.45\linewidth}
    \includegraphics[width=\linewidth]{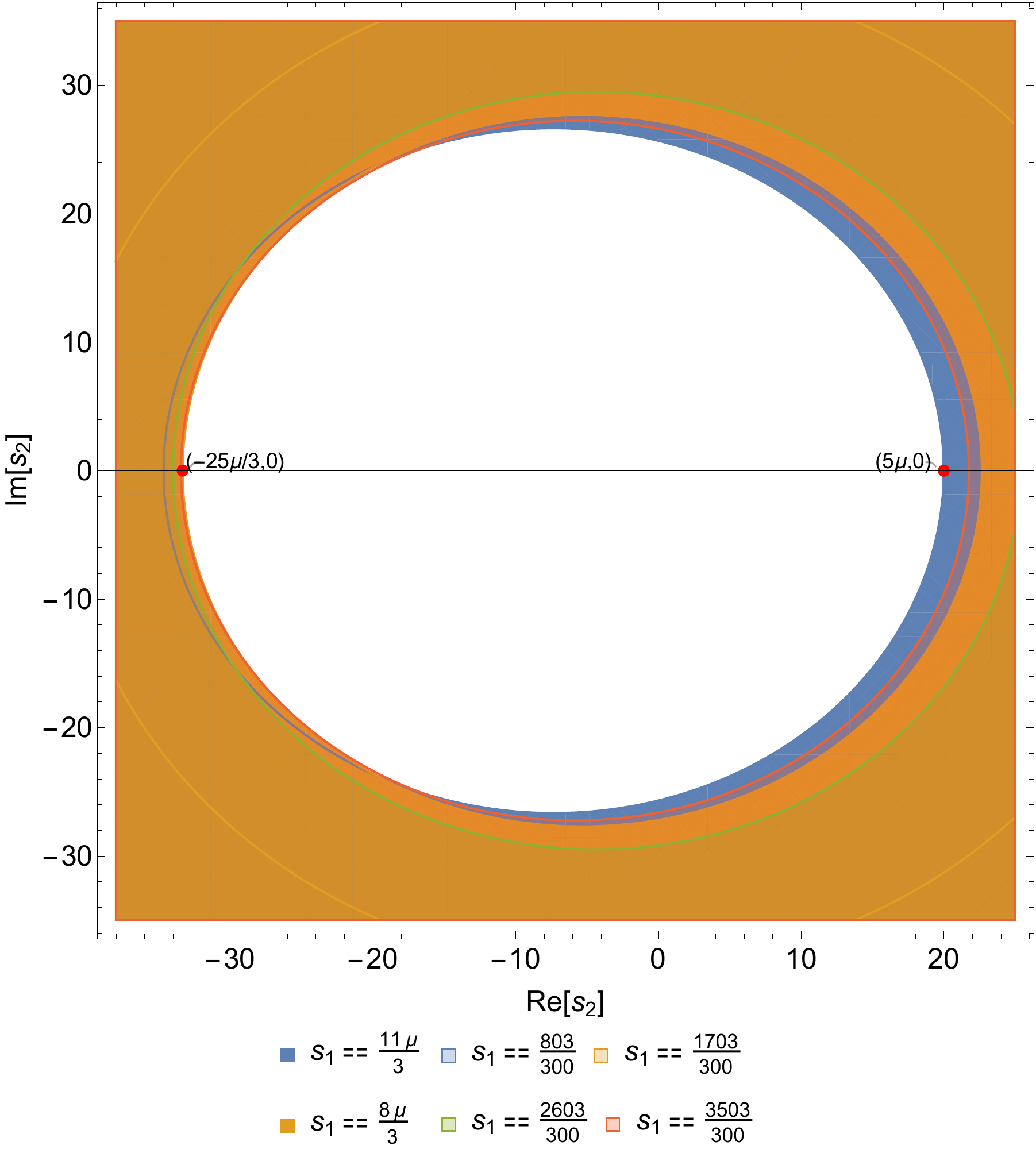}
    \caption{The analyticity domain in complex $s_2$-plane for various $s_1$ in the range $\frac{2 \mu }{3}<s_1<\frac{11 \mu }{3}$.}
\label{fig:martin_region1}
  \end{subfigure}
  \begin{subfigure}[b]{0.45\linewidth}
    \includegraphics[width=\linewidth]{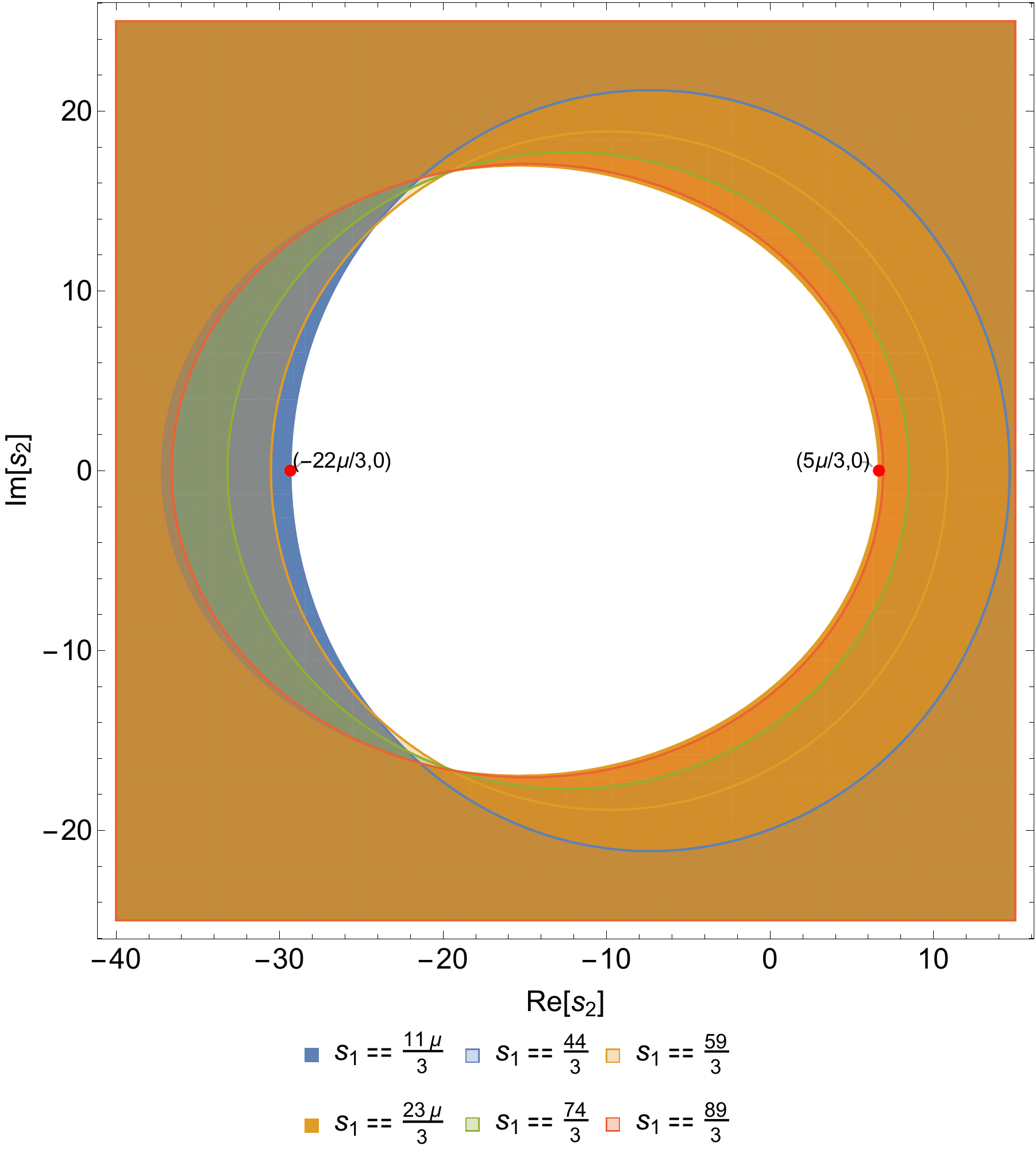}
    \caption{The analyticity domain in complex $s_2$-plane for various $s_1$ in the range $\frac{11 \mu }{3}<s_1<\frac{23 \mu }{3}$.}
\label{fig:martin_region2}
  \end{subfigure}
  \begin{subfigure}[b]{0.45\linewidth}
    \includegraphics[width=\linewidth]{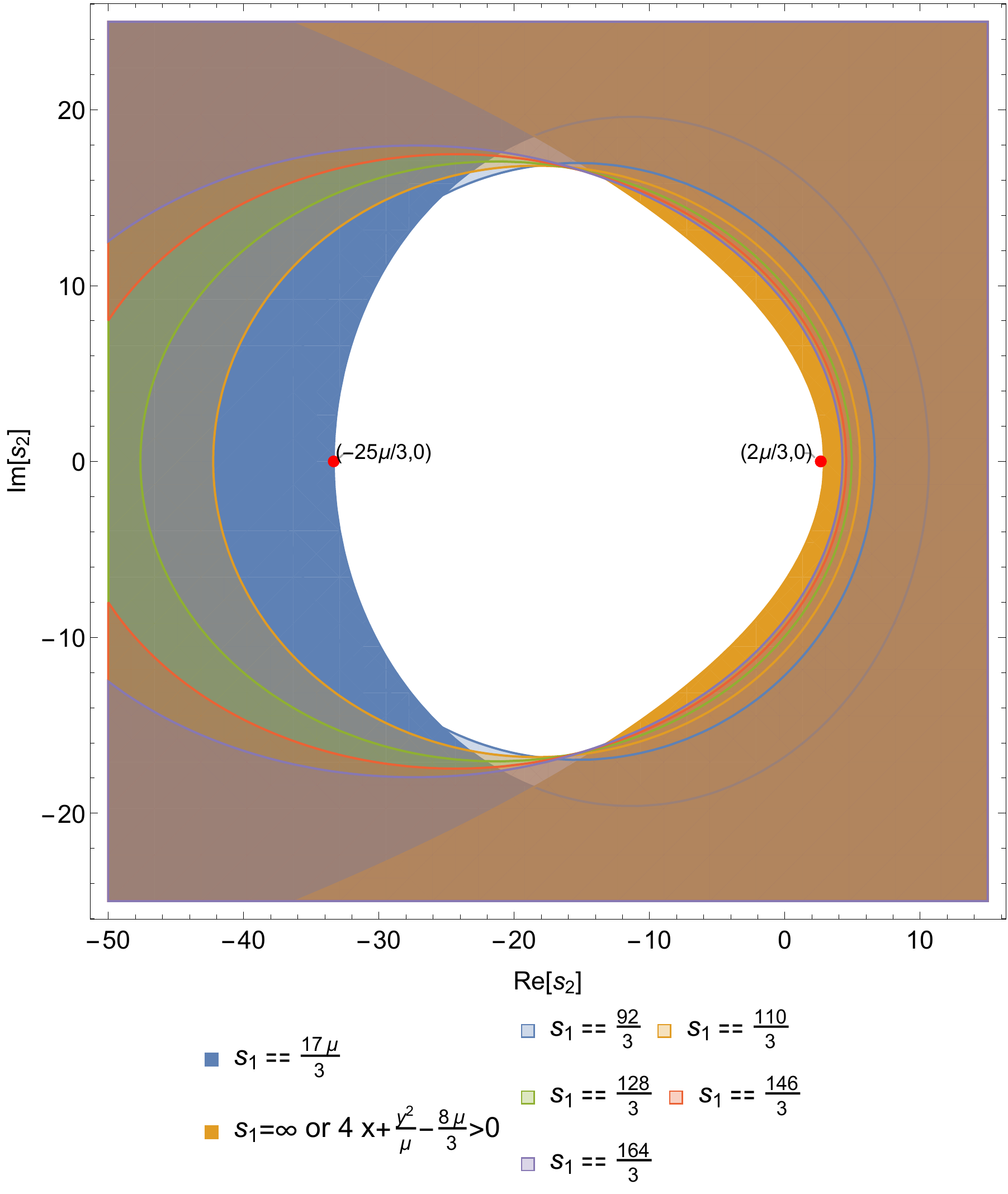}
   \caption{The analyticity domain in complex $s_2$-plane for various $s_1$ in the range $\frac{23 \mu }{3}<s_1<\infty$.}
\label{fig:martin_region3}
  \end{subfigure}
  \begin{subfigure}[b]{0.45\linewidth}
    \includegraphics[width=\linewidth]{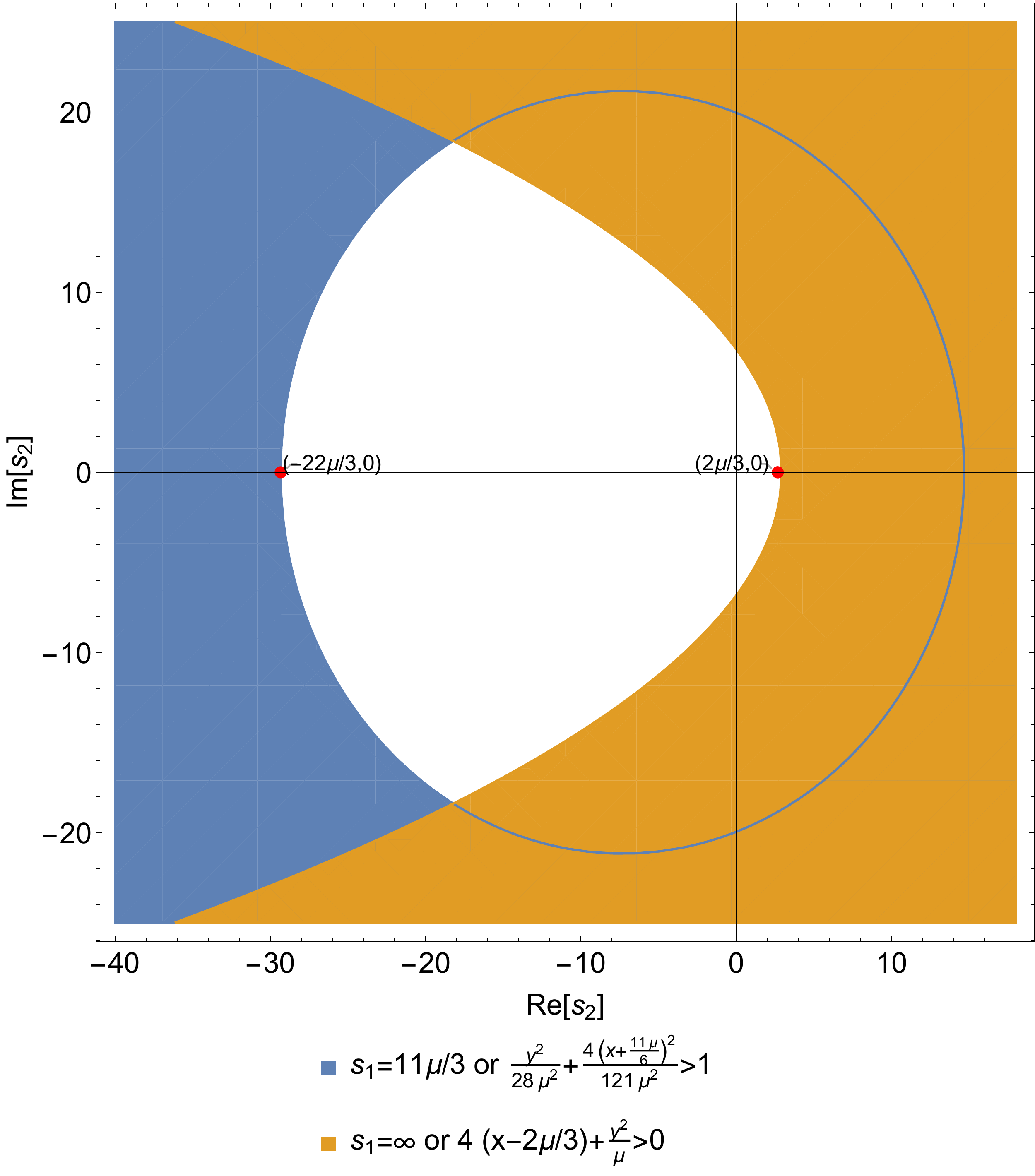}
   \caption{The analyticity domain in complex $s_2$-plane for various $s_1$ in the range $\frac{2 \mu }{3}<s_1<\infty$. }
\label{fig:martin_region_all}
  \end{subfigure}
  \caption{Lehmann-Martin ellipses. Shaded region is excluded.}
  \label{fig:all4martin}
\end{figure}

\subsubsection*{Region 1}


Ellipses for various values of $s_1$ for $LM_1(s_1)$ is shown \eqref{fig:martin_region1}. The dominant ellipses are: the ellipse corresponds to  $s_1=11\m/3$ which touches the real axis at $s_2=5\m$ indicated in the figure and the ellipse that minimises the intersection of the ellipse in the negative  real axis, which can be found following way. Note that negative axis it touches when $\Im(s_2)=0$, which gives 
\be
\Re(s_2)=\frac{-14 \mu ^2-27 \mu  s_1-9 s_1^2}{3 \left(3 s_1-2 \mu \right)}
\ee
The value of $s_1= 8\m/3$ minimizes the negative real axis intercept, which is $\Re(s_2)=-\frac{25\m}{3}$, indicated in the figure \eqref{fig:martin_region1}.

\subsubsection*{Region 2}
Ellipses for various values of $s_1$ for $LM_2(s_1)$ is shown \eqref{fig:martin_region2}
The dominant ellipses are: the ellipse corresponds to  $s_1=23\m/3$ which touches the real axis at $s_2=5\m/3$ indicated in the figure and the ellipse that minimises the intersect in negative axis which can be found following way. Note that negative axis it touches when $\Im(s_2)=0$, which gives 
\be
\Re(s_2)=\frac{143 \mu ^2-3 \mu  s_1}{3 \left(\mu +3 s_1\right)}
\ee
The value of $s_1= \frac{11 \mu }{3}$ minimizes the negative axis intercept, which is $\Re(s_2)=-\frac{22\m}{3}$, indicated in the figure \eqref{fig:martin_region2}.

\subsubsection*{Region 3}
Ellipses for various values of $s_1$ for $LM_3(s_1)$ is shown \eqref{fig:martin_region3}.
The dominant ellipses are: the ellipse corresponds to  $s_1\to\infty$ which touches the real axis at $s_2=2\m/3$ indicated in the figure and the ellipse that minimises the intersect in negative axis which can be found following way. Note that negative axis it touches when $\Im(s_2)=0$, which gives 
\be
\Re(s_2)=\frac{-14 \mu ^2+27 \mu  s_1-9 s_1^2}{3 \left(3 s_1-11 \mu \right)}
\ee
The value of $s_1= \frac{17 \mu }{3}$ minimizes the negative axis intercept, which is $\Re(s_2)=-\frac{25\m}{3}$, indicated in the figure \eqref{fig:martin_region3}.

\subsubsection*{Overall region}

The overall domain traced out by all three $LM_1(s_1), LM_2(s_1),LM_3(s_1)$ is shown \eqref{fig:martin_region_all}:
\begin{itemize}
\item The value of $s_1= \frac{11 \mu }{3}$ minimizes the negative axis intercept, which is $\Re(s_2)=-\frac{22\m}{3}$ for the ellipse $LM_2(s_1)$. The exact equation is given by
\be
\frac{4 \left(\frac{11 \mu }{6}+\Re(s_2)\right)^2}{121 \mu ^2}+\frac{\Im(s_2)^2}{28 \mu ^2}=1
\ee
\item The ellipse corresponds to  $s_1\to\infty$ which touches the real axis at $s_2=2\m/3$ or $LM_3(s_1=\infty)$.
\be 
4\left(\Re(s_2)-\frac{2\mu}{3}\right)+\frac{\Im(s_2)^2}{\mu}=0
\ee

\end{itemize}


\begin{thebibliography}{99}
\bibliographystyle{utphys}



\bibitem{Dispcause}
S.~Mandelstam,
``Determination of the pion - nucleon scattering amplitude from dispersion relations and unitarity. General theory,''
Phys. Rev. \textbf{112}, 1344-1360 (1958)\\
H.~M~Nussenzveig, ``Causality and Dispersion Relations,'' Academic Press, 1972.

   
\bibitem{Martinlec}
A.~Martin,
``Scattering Theory: Unitarity, Analyticity and Crossing,''
Lect. Notes Phys. \textbf{3}, 1-117 (1969)\\

\bibitem{Froissartbnd}
M.~Froissart,
``Asymptotic behavior and subtractions in the Mandelstam representation,''
Phys. Rev. \textbf{123}, 1053-1057 (1961)\\
P.~Haldar and A.~Sinha,
``Froissart bound for/from CFT Mellin amplitudes,''
SciPost Phys. \textbf{8}, 095 (2020)
[arXiv:1911.05974 [hep-th]].




\bibitem{Meltzer}
D.~Meltzer,
``Dispersion Formulas in QFTs, CFTs, and Holography,''
JHEP \textbf{05}, 098 (2021)
[arXiv:2103.15839 [hep-th]].\\
D.~Meltzer,
``The Inflationary Wavefunction from Analyticity and Factorization,''
JCAP \textbf{12} (2021) no.12, 018
[arXiv:2107.10266 [hep-th]].



   \bibitem{nima}
A.~Adams, N.~Arkani-Hamed, S.~Dubovsky, A.~Nicolis and R.~Rattazzi,
``Causality, analyticity and an IR obstruction to UV completion,''
JHEP \textbf{10}, 014 (2006)
[arXiv:hep-th/0602178 [hep-th]].


\bibitem{Mizera}
H.~S.~Hannesdottir and S.~Mizera,
``What is the $i\varepsilon$ for the S-matrix?,''
[arXiv:2204.02988 [hep-th]].



\bibitem{RMTZ}
C.~de Rham, S.~Melville, A.~J.~Tolley and S.~Y.~Zhou,
``Positivity bounds for scalar field theories,''
Phys. Rev. D \textbf{96}, no.8, 081702 (2017)
[arXiv:1702.06134 [hep-th]].


   \bibitem{rattazzi}
B.~Bellazzini, J.~Elias Mir\'o, R.~Rattazzi, M.~Riembau and F.~Riva,
``Positive Moments for Scattering Amplitudes,''
Phys. Rev. D \textbf{104} (2021) no.3, 036006
[arXiv:2011.00037 [hep-th]].
   
\bibitem{TWZ}
A.~J.~Tolley, Z.~Y.~Wang and S.~Y.~Zhou,
``New positivity bounds from full crossing symmetry,''
JHEP \textbf{05} (2021), 255
[arXiv:2011.02400 [hep-th]].

\bibitem{Wang:2020jxr}
Y.~J.~Wang, F.~K.~Guo, C.~Zhang and S.~Y.~Zhou,
``Generalized positivity bounds on chiral perturbation theory,''
JHEP \textbf{07}, 214 (2020)
[arXiv:2004.03992 [hep-ph]].




\bibitem{chout}
S.~Caron-Huot and V.~Van Duong,
``Extremal Effective Field Theories,''
JHEP \textbf{05} (2021), 280
[arXiv:2011.02957 [hep-th]].

\bibitem{chout2}
S.~Caron-Huot, D.~Mazac, L.~Rastelli and D.~Simmons-Duffin,
``Sharp Boundaries for the Swampland,''
JHEP \textbf{07} (2021), 110
[arXiv:2102.08951 [hep-th]].

S.~Caron-Huot, Y.~Z.~Li, J.~Parra-Martinez and D.~Simmons-Duffin,
``Causality constraints on corrections to Einstein gravity,''
[arXiv:2201.06602 [hep-th]].\\
S.~Caron-Huot, Y.~Z.~Li, J.~Parra-Martinez and D.~Simmons-Duffin,
``Graviton partial waves and causality in higher dimensions,''
[arXiv:2205.01495 [hep-th]].

\bibitem{chout3}
S.~Caron-Huot, D.~Mazac, L.~Rastelli and D.~Simmons-Duffin,
``AdS Bulk Locality from Sharp CFT Bounds,''
JHEP \textbf{11} (2021), 164
[arXiv:2106.10274 [hep-th]].

\bibitem{Vichi}
J.~Henriksson, B.~McPeak, F.~Russo and A.~Vichi,
``Rigorous Bounds on Light-by-Light Scattering,''
[arXiv:2107.13009 [hep-th]].\\
J.~Henriksson, B.~McPeak, F.~Russo and A.~Vichi,
``Bounding Violations of the Weak Gravity Conjecture,''
[arXiv:2203.08164 [hep-th]].


\bibitem{nimayutin}
N. Arkani-Hamed, Y.-T. Huang
``Lectures at the CERN winter school on supergravity, strings and gauge theory.''

\bibitem{green}
M.~B.~Green and C.~Wen,
``Superstring amplitudes, unitarily, and Hankel determinants of multiple zeta values,''
JHEP \textbf{11}, 079 (2019)
[arXiv:1908.08426 [hep-th]].
\bibitem{nimayutin2}
N.~Arkani-Hamed, T.~C.~Huang and Y.~T.~Huang,
``The EFT-Hedron,''
JHEP \textbf{05}, 259 (2021)
[arXiv:2012.15849 [hep-th]].
\bibitem{yutin2}
L.~Y.~Chiang, Y.~t.~Huang, W.~Li, L.~Rodina and H.~C.~Weng,
``Into the EFThedron and UV constraints from IR consistency,''
JHEP \textbf{03} (2022), 063
[arXiv:2105.02862 [hep-th]].

\bibitem{sasha}
Z.~Bern, D.~Kosmopoulos and A.~Zhiboedov,
``Gravitational Effective Field Theory Islands, Low-Spin Dominance, and the Four-Graviton Amplitude,''
J. Phys. A \textbf{54} (2021) no.34, 344002
[arXiv:2103.12728 [hep-th]].\\
K.~H\"aring and A.~Zhiboedov,
``Gravitational Regge bounds,''
[arXiv:2202.08280 [hep-th]].






\bibitem{Davis:2021oce}
A.~C.~Davis and S.~Melville,
``Scalar Fields Near Compact Objects: Resummation versus UV Completion,''
JCAP \textbf{11} (2021), 012
[arXiv:2107.00010 [gr-qc]].



\bibitem{PHASAZ}
P.~Haldar, A.~Sinha and A.~Zahed,
``Quantum field theory and the Bieberbach conjecture,''
SciPost Phys. \textbf{11}, 002 (2021)
[arXiv:2103.12108 [hep-th]].

\bibitem{PRAS}
P.~Raman and A.~Sinha,
``QFT, EFT and GFT,''
JHEP \textbf{12} (2021), 203
[arXiv:2107.06559 [hep-th]].

\bibitem{AZ}
A.~Zahed,
``Positivity and geometric function theory constraints on pion scattering,''
JHEP \textbf{12} (2021), 036
[arXiv:2108.10355 [hep-th]].

\bibitem{AS}
A.~Sinha,
``Dispersion relations and knot theory,''
[arXiv:2204.13986 [hep-th]].



\bibitem{joaopaper}
J.~Penedones, J.~A.~Silva and A.~Zhiboedov,
``Nonperturbative Mellin Amplitudes: Existence, Properties, Applications,''
JHEP \textbf{08}, 031 (2020)
[arXiv:1912.11100 [hep-th]].
\bibitem{joaopaper2}
D.~Carmi, J.~Penedones, J.~A.~Silva, and A.~Zhiboedov,
``Applications of dispersive sum rules: $\epsilon$-expansion and holography,''
SciPost Phys. \textbf{10} (2021) no.6, 145
[arXiv:2009.13506 [hep-th]].
   
   
 \bibitem{cmrs}
S.~Caron-Huot, D.~Mazac, L.~Rastelli and D.~Simmons-Duffin,
``Dispersive CFT Sum Rules,''
JHEP \textbf{05} (2021), 243
[arXiv:2008.04931 [hep-th]]. 




   \bibitem{RGASAZ}
R.~Gopakumar, A.~Sinha and A.~Zahed,
``Crossing Symmetric Dispersion Relations for Mellin Amplitudes,''
Phys. Rev. Lett. \textbf{126}, no.21, 211602 (2021)
[arXiv:2101.09017 [hep-th]].


   
\bibitem{AK}
G.~Auberson and N.~N.~Khuri,
``Rigorous parametric dispersion representation with three-channel symmetry,''
Phys. Rev. D \textbf{6}, 2953-2966 (1972)












\bibitem{ASAZ}
A.~Sinha and A.~Zahed,
``Crossing Symmetric Dispersion Relations in Quantum Field Theories,''
Phys. Rev. Lett. \textbf{126}, no.18, 181601 (2021)
[arXiv:2012.04877 [hep-th]].




\bibitem{roy}
G.~Mahoux, S.~M.~Roy and G.~Wanders,
``Physical pion pion partial-wave equations based on three channel crossing symmetry,''
Nucl. Phys. B \textbf{70}, 297-316 (1974)

\bibitem{Roskies:1970uj}
R.~Roskies,
``Crossing restrictions on pi pi partial waves,''
Nuovo Cim. A \textbf{65}, 467-490 (1970)


 \bibitem{anant}
B.~Ananthanarayan,
``The Low-energy expansion for pion pion scattering and crossing symmetry in dispersion relations,''
Phys. Rev. D \textbf{58}, 036002 (1998)
[arXiv:hep-ph/9802338 [hep-ph]].






%

\bibitem{usprl}
R.~Gopakumar, A.~Kaviraj, K.~Sen and A.~Sinha,
``Conformal Bootstrap in Mellin Space,''
Phys. Rev. Lett. \textbf{118}, no.8, 081601 (2017)
[arXiv:1609.00572 [hep-th]].\\
R.~Gopakumar, A.~Kaviraj, K.~Sen and A.~Sinha,
``A Mellin space approach to the conformal bootstrap,''
JHEP \textbf{05}, 027 (2017)
[arXiv:1611.08407 [hep-th]].\\
R.~Gopakumar and A.~Sinha,
``On the Polyakov-Mellin bootstrap,''
JHEP \textbf{12}, 040 (2018)
[arXiv:1809.10975 [hep-th]].

\bibitem{WhitaWat}
E.T.~Whittaker and G.N.~Watson,
``A Course of Modern Analysis,''
Cambridge University Press, 1996.
   \bibitem{Lehmann1}
   H.~Lehmann,
   ``Analytic properties of scattering amplitudes as functions of momentum transfer,''
   Nuovo Cim. \textbf{10}, no.4, 579-589 (1958)
   
   
   \bibitem{MartinExt1}
   A.~Martin,
   ``Extension of the axiomatic analyticity domain of scattering amplitudes by unitarity. 1.,''
   Nuovo Cim. A \textbf{42}, 930-953 (1965).
   \bibitem{BELG}
   J.~Bros, H.~Epstein and V.~J.~Glaser,
   ``Some rigorous analyticity properties of the four-point function in momentum space,''
   Nuovo Cim. \textbf{31}, 1265-1302 (1964),\\
   H.~Lehmann,
   ``Analytic properties of scattering amplitudes as functions of momentum transfer,''
   Nuovo Cim. \textbf{10}, no.4, 579-589 (1958).
   
   \bibitem{MartinLec}
   A.~Martin,
   `` Scattering Theory: Unitarity, Analyticity and Crossing,''
   Lect. Notes Phys. \textbf{3} (1969).
%

   \bibitem{MartinExt2}
   A.~Martin,
   ``Extension of the axiomatic analyticity domain of scattering amplitudes by unitarity.-II.,''
   Nuovo Cim. A \textbf{44}, 1219-1244 (1966).
   
\bibitem{ABAS}
A.~Bose, A.~Sinha and S.~S.~Tiwari,
``Selection rules for the S-Matrix bootstrap,''
SciPost Phys. \textbf{10} (2021) no.5, 122
[arXiv:2011.07944 [hep-th]].



\bibitem{ABPHAS}
A.~Bose, P.~Haldar, A.~Sinha, P.~Sinha and S.~S.~Tiwari,
``Relative entropy in scattering and the S-matrix bootstrap,''
SciPost Phys. \textbf{9}, 081 (2020)
[arXiv:2006.12213 [hep-th]].





\bibitem{Pol}
A.~M.~Polyakov,
``Nonhamiltonian approach to conformal quantum field theory,''
Zh. Eksp. Teor. Fiz. \textbf{66}, 23-42 (1974)

\bibitem{ks}
K.~Sen and A.~Sinha,
``On critical exponents without Feynman diagrams,''
J. Phys. A \textbf{49}, no.44, 445401 (2016)
[arXiv:1510.07770 [hep-th]].


    
\bibitem{ON}
P.~Dey, A.~Kaviraj and A.~Sinha,
``Mellin space bootstrap for global symmetry,''
JHEP \textbf{07}, 019 (2017)
[arXiv:1612.05032 [hep-th]].

\bibitem{PFKGASAZ}
P.~Ferrero, K.~Ghosh, A.~Sinha and A.~Zahed,
``Crossing symmetry, transcendentality and the Regge behaviour of 1d CFTs,''
JHEP \textbf{07}, 170 (2020)
[arXiv:1911.12388 [hep-th]].

%

\bibitem{Ghosh:2021ruh}
K.~Ghosh, A.~Kaviraj and M.~F.~Paulos,
``Charging Up the Functional Bootstrap,''
JHEP \textbf{10} (2021), 116
[arXiv:2107.00041 [hep-th]].


\bibitem{Apratim}
A.~Kaviraj,
``Crossing antisymmetric Polyakov blocks + Dispersion relation,''
JHEP \textbf{01} (2022), 005
[arXiv:2109.02658 [hep-th]].

\bibitem{Alday}
L.~F.~Alday, T.~Hansen and J.~A.~Silva,
``AdS Virasoro-Shapiro from dispersive sum rules,''
[arXiv:2204.07542 [hep-th]].

\bibitem{Kundu:2021qpi}
S.~Kundu,
``Swampland Conditions for Higher Derivative Couplings from CFT,''
JHEP \textbf{01} (2022), 176
[arXiv:2104.11238 [hep-th]].




\bibitem{Paulos:2020zxx}
M.~F.~Paulos,
``Dispersion relations and exact bounds on CFT correlators,''
JHEP \textbf{08} (2021), 166
[arXiv:2012.10454 [hep-th]].\\
L.~C\'ordova, Y.~He and M.~F.~Paulos,
``From conformal correlators to analytic S-matrices: CFT$_1$/QFT$_2$,''
[arXiv:2203.10840 [hep-th]].


\bibitem{positivegeo}
N.~Arkani-Hamed, Y.~T.~Huang and S.~H.~Shao,
``On the Positive Geometry of Conformal Field Theory,''
JHEP \textbf{06}, 124 (2019)
[arXiv:1812.07739 [hep-th]].\\
K.~Sen, A.~Sinha and A.~Zahed,
``Positive geometry in the diagonal limit of the conformal bootstrap,''
JHEP \textbf{11}, 059 (2019)
[arXiv:1906.07202 [hep-th]].\\
Y.~T.~Huang, W.~Li and G.~L.~Lin,
``The geometry of optimal functionals,''
[arXiv:1912.01273 [hep-th]].



\bibitem{SGASPR}
S.~Ghosh, A.~Sinha and P.~Raman,
``Celestial insights into the S-matrix bootstrap,''
[arXiv:2204.07617 [hep-th]].

 \bibitem{andrea}
A.~L.~Guerrieri, J.~Penedones and P.~Vieira,
``Bootstrapping QCD Using Pion Scattering Amplitudes,''
Phys. Rev. Lett. \textbf{122}, no.24, 241604 (2019)
[arXiv:1810.12849 [hep-th]].
   
   
\bibitem{dual}
A.~L.~Guerrieri, A.~Homrich and P.~Vieira,
``Dual S-matrix bootstrap. Part I. 2D theory,''
JHEP \textbf{11}, 084 (2020)
[arXiv:2008.02770 [hep-th]].\\
Y.~He and M.~Kruczenski,
``S-matrix bootstrap in 3+1 dimensions: regularization and dual convex problem,''
JHEP \textbf{08} (2021), 125
[arXiv:2103.11484 [hep-th]].\\
J.~E.~Mir\'o and A.~Guerrieri,
``Dual EFT Bootstrap: QCD flux tubes,''
JHEP \textbf{10} (2021), 126
[arXiv:2106.07957 [hep-th]].\\
A.~Guerrieri and A.~Sever,
``Rigorous bounds on the Analytic $S$-matrix,''
Phys. Rev. Lett. \textbf{127} (2021) no.25, 251601
[arXiv:2106.10257 [hep-th]].

 
\bibitem{spinning}
S.~D.~Chowdhury, K.~Ghosh, P.~Haldar, P.~Raman and A.~Sinha,
``Crossing Symmetric Spinning S-matrix Bootstrap: EFT bounds,''
[arXiv:2112.11755 [hep-th]].

J.~Davighi, S.~Melville and T.~You,
``Natural selection rules: new positivity bounds for massive spinning particles,''
JHEP \textbf{02} (2022), 167
[arXiv:2108.06334 [hep-th]].


C.~Zhang,
``SMEFTs living on the edge: determining the UV theories from positivity and extremality,''
[arXiv:2112.11665 [hep-ph]].

\bibitem{Bern:2022yes}
Z.~Bern, E.~Herrmann, D.~Kosmopoulos and R.~Roiban,
``Effective Field Theory Islands from Perturbative and Nonperturbative Four-Graviton Amplitudes,''
[arXiv:2205.01655 [hep-th]].

\bibitem{Multi}
Z.~Z.~Du, C.~Zhang and S.~Y.~Zhou,
``Triple crossing positivity bounds for multi-field theories,''
JHEP \textbf{12} (2021), 115
[arXiv:2111.01169 [hep-th]].

B.~Alvarez, J.~Bijnens and M.~Sj\"o,
``NNLO positivity bounds on chiral perturbation theory for a general number of flavours,''
JHEP \textbf{03} (2022), 159
[arXiv:2112.04253 [hep-ph]].

\bibitem{Rastelli}
J.~Albert and L.~Rastelli,
``Bootstrapping Pions at Large $N$,''
[arXiv:2203.11950 [hep-th]].




\bibitem{Karateev:2022jdb}
D.~Karateev, J.~Marucha, J.~Penedones and B.~Sahoo,
``Bootstrapping the $a$-anomaly in $4d$ QFTs,''
[arXiv:2204.01786 [hep-th]].




\bibitem{Moments}
X.~Li, K.~Mimasu, K.~Yamashita, C.~Yang, C.~Zhang and S.~Y.~Zhou,
``Moments for positivity: using Drell-Yan data to test positivity bounds and reverse-engineer new physics,''
[arXiv:2204.13121 [hep-ph]].




\bibitem{Causality}
C.~Y.~R.~Chen, C.~de Rham, A.~Margalit and A.~J.~Tolley,
``A cautionary case of casual causality,''
JHEP \textbf{03} (2022), 025
[arXiv:2112.05031 [hep-th]].\\
C.~de Rham, A.~J.~Tolley and J.~Zhang,
``Causality Constraints on Gravitational Effective Field Theories,''
Phys. Rev. Lett. \textbf{128} (2022) no.13, 131102
doi:10.1103/PhysRevLett.128.131102
[arXiv:2112.05054 [gr-qc]].


\bibitem{Reverse}
L.~Alberte, C.~de Rham, S.~Jaitly and A.~J.~Tolley,
``Reverse Bootstrapping: IR Lessons for UV Physics,''
Phys. Rev. Lett. \textbf{128} (2022) no.5, 5
[arXiv:2111.09226 [hep-th]].


\bibitem{Yutinnew1}
L.~Y.~Chiang, Y.~t.~Huang, L.~Rodina and H.~C.~Weng,
``De-projecting the EFThedron,''
[arXiv:2204.07140 [hep-th]].








\bibitem{Hor}
Lars~H\"{o}rmander, ``An Introduction to Complex Analysis in Several Variable'', North-Holland, Ellsevier (1990).

   
%
%
%
%
%









%
%






%
%
%




\end{thebibliography}
\end{document}